\documentclass[apj]{emulateapj}
\shorttitle{Near- and Mid-IR Spectroscopy of Gravitationally Lensed
  Star-Forming Galaxies} 
\shortauthors{Rujopakarn et al.}

\newcommand{\LMIPS}{\mbox{$L(24~\micron)$}}
\newcommand{\Lsun}{\mbox{$L_\odot$}}
\newcommand{\LTIR}{\mbox{$L({\rm TIR})$}} 
\newcommand{\LIRSD}{\mbox{$\Sigma_{L({\rm TIR})}$}} 
\newcommand{\Msun}{\mbox{${\rm M}_\odot$}}
 
\newcommand{\Av}{\mbox{A$_{\rm v}$}} 
\newcommand{\AvIR}{\mbox{A$^{{\rm IR}}_{\rm v}$}} 

\slugcomment{Accepted for publication in The Astrophysical Journal, June 18, 2012}
\begin{document}
\title{LBT and {\it Spitzer} Spectroscopy of Star-Forming Galaxies at
  1 $< z <$ 3: Extinction and Star Formation Rate Indicators}

\author{W. Rujopakarn\altaffilmark{1}, G. H. Rieke\altaffilmark{1},
  C. J. Papovich\altaffilmark{2}, B. J. Weiner\altaffilmark{1},
  J. R. Rigby\altaffilmark{3}, M. Rex\altaffilmark{1},
  F. Bian\altaffilmark{1}, O. P. Kuhn\altaffilmark{4},
  D. Thompson\altaffilmark{4}}
\altaffiltext{1}{Steward Observatory, The University of Arizona,
  Tucson, AZ 85721, USA; wiphu@as.arizona.edu} 
\altaffiltext{2}{George P. and Cynthia Woods Mitchell Institute for
  Fundamental Physics and Astronomy, Department of Physics and
  Astronomy, Texas A\&M University, College Station, TX 77843, USA}
\altaffiltext{3}{Observational Cosmology Lab, NASA Goddard Space
  Flight Center, Greenbelt, MD 20771, USA}
\altaffiltext{4}{Large Binocular Telescope Observatory, The University
  of Arizona, Tucson, AZ 85721, USA} 

\begin{abstract}
We present spectroscopic observations in the rest-frame optical and
near- to mid-infrared wavelengths of four gravitationally lensed
infrared (IR) luminous star-forming galaxies at redshift 1 $< z <$ 3 
from the LUCIFER instrument on the Large Binocular Telescope and
the Infrared Spectrograph on {\it Spitzer}. The sample was
selected to represent pure, actively star-forming systems, absent of
active galactic nuclei. The large lensing magnifications result
in high signal-to-noise spectra that can probe faint IR recombination
lines, including Pa$\alpha$ and Br$\alpha$ at high redshifts. The
sample was augmented by three lensed galaxies with similar suites of
unpublished data and observations from the literature, resulting in
the final sample of seven galaxies. We use the IR recombination lines
in conjunction with H$\alpha$ observations to probe the extinction,
\Av, of these systems, as well as testing star formation rate (SFR)
indicators against the SFR measured by fitting spectral energy
distributions to far-IR photometry. Our galaxies occupy a range of
\Av\ from $\sim 0$ to 5.9 mag, larger than previously known for a
similar range of IR luminosities at these redshifts. Thus, estimates
of SFR even at $z \sim 2$ must take careful count of extinction in the
most IR luminous galaxies. We also measure extinction by comparing SFR
estimates from optical emission lines with those from far-IR
measurements. The comparison of results from these two independent
methods indicates a large variety of dust distribution scenarios at $1
< z < 3$. Without correcting for dust extinction, the H$\alpha$ SFR
indicator underestimates the SFR; the size of the necessary correction
depends on the IR luminosity and dust distribution
scenario. Individual SFR estimates based on the 6.2 $\micron$ PAH
emission line luminosity do not show a systematic discrepancy with
extinction, although a considerable, $\sim$ 0.2 dex scatter is
observed.
\end{abstract}

\section{INTRODUCTION}
The evolution of the star formation rate (SFR) of galaxies is a
central topic to the study of galaxy evolution. It is generally 
agreed that the SFR density of the Universe has declined by an order
of magnitude since $z \sim 1$ to the present \citep[e.g.,][]{LeF05,
  Rujopakarn10, Magnelli11}. The exact epoch of the peak of the SFR
history is not known precisely, although it appears to be constrained
to be within $1 < z < 3$ \citep[e.g.,][]{PPG05, Reddy08, Rodighiero10,
  Magnelli11}.

At this redshift range, the primary SFR indicators are based on
infrared (IR), optical, and extinction-corrected ultraviolet (UV)
observations; the resulting SFR estimates commonly disagree with
each other by more than a factor of two \citep[e.g.,][]{Reddy08}. The
majority of star formation at these redshifts is known to occur in
optically extincted star-forming regions in IR luminous galaxies
\citep[e.g.,][]{LeF05, Dole06, Berta11}, and the uncertainties
resulting from the extinction undermine our ability to understand the
SFR history during this era. A better understanding of the SFR history
will have important implications on the cosmic stellar mass build-up
and metal production. For example, there is a significant discrepancy
between the expected metal abundance derived from the SFR history and
the observed abundance \citep{Bouche07}, with some studies indicating
the difference to be nearly an order of magnitude, and as a result
placing the peak of the SFR history as far back as $z \sim 4$
\citep{Kobayashi07}.

The situation necessitates more efforts into exploring the nature of
optical extinction in high redshift galaxies using unbiased measures
as well as studying the effect of extinction on various SFR indicators.

Extinction measurements based on optical emission lines could be
biased in highly obscured environments because the indicator  
only probes the outer layer of the star-forming regions, where the
extinction is relatively low. This effect is observed locally
\citep[e.g.,][]{AH06}, but only because observations are available
for longer 
wavelength IR recombination lines (e.g. Pa$\alpha$ and Br$\alpha$)
that are less affected by extinction. In this work, we extend this
technique out to $z = 3$ by comparing the strength of the H$\alpha$
line with those of Pa$\alpha$ and Br$\alpha$. Since the latter are in
wavelength regions with  $10-20$ times less extinction than H$\alpha$,
they provide a measurement of extinction through the entire
star-forming region. With an unbiased estimate of extinction, we can 
address its effects on SFR indicators in this critical redshift range.

In this paper, we study a wide range of star formation diagnostics in
seven gravitationally lensed star-forming galaxies at $1 < z < 3$. We
observed four of these galaxies spectroscopically with the LUCIFER1
near-IR imaging spectrograph on the Large Binocular Telescope (LBT) to
measure the H$\alpha$ line flux in the near-IR, and with the Infrared
Spectrograph (IRS, $5-38$ $\micron$) on the {\it Spitzer} Space
Telescope to observe the wavelength regions covering Pa$\alpha$ and/or
Br$\alpha$ lines as well as aromatic emission lines and emission
complexes (commonly attributed to, and hereafter, polycyclic aromatic
hydrocarbons or PAH). The sample comprises Abell 2218b, Abell 2667a,
Abell 2218a, and Abell 1835a at redshift 1.03, 1.03, 2.52, and 2.57,
respectively (magnification 6$\times$ - 27$\times$). The IRS
observations at longer wavelengths are further used to compare SFR
estimates from the 6.2 $\micron$ PAH feature. They were carried out
under {\it Spitzer} program ID 82, 30775, 50586 (PI G. Rieke); and
40443, and 50372 (PI C. Papovich). Our sample is augmented by three
galaxies with similar suites of observations from unpublished data and
the literature: SDSSJ120601+5142 (hereafter the Clone), the Lyman
break galaxy LBG MS 1512-cB58 (hereafter cB58), and the 8 O'clock arc,
at redshift 2.00, 2.73, and 2.73, respectively. Our galaxies were also 
observed with the MIPS instrument \citep{Rieke04} on {\it Spitzer} at
24 and 70 $\micron$ by \citet{Rigby08}, allowing us to combine the
MIPS data with far-IR photometry from the literature to estimate SFR
via \LTIR\ obtained by fitting the spectral energy distribution (SED)
across the peak of the far-IR emission. In addition, we used the 24
$\micron$ monochromatic indicator \citep{Rujopakarn12} to estimate \LTIR. 

This paper is organized as follow. We discuss the LBT and {\it
  Spitzer} observations and data reduction in Section \ref{sec:obs}; 
extinction, metallicity, \LTIR, and SFR measurements in Section
\ref{sec:result}; compare SFR indicators in Section
\ref{sec:discuss}; and make concluding remarks in Section
\ref{sec:conclusions}. We assume a $\Lambda$CDM cosmology with
$\Omega_m = 0.3$, $\Omega_{\Lambda} = 0.7$, and $H_0 =
70$~km~s$^{-1}$Mpc$^{-1}$ throughout this paper. 

\section{OBSERVATIONS AND DATA REDUCTION}\label{sec:obs} 
Here we describe the selection of our sample (Section
\ref{sec:obs_sample}); the spectroscopic observations with LBT/LUCIFER 
along with the near-IR data reduction (Section \ref{sec:obs_LBT} and
Section \ref{sec:obs_LBT_reduc}) and with {\it Spitzer} (Section
\ref{sec:obs_IRS}); data compilation from the literature (Section 
\ref{sec:obs_lit}); additional photometric observations with {\it
  Spitzer} MIPS and the respective data reduction procedure (Section
\ref{sec:obs_MIPS}) that results in the final data set for our sample
of galaxies. 

\begin{center}
\begin{deluxetable*}{lccccc}
\tablewidth{0pt}
\tablecaption{The Sample of High-Redshift Strongly-Lensed Star-Forming
  Galaxies} 
\tablehead{Source & $z$ & R.A. & Decl. & Lensing &
  Reference \\
& & (J2000) & (J2000) & Magnification &}
\startdata
Abell 2218b & 1.034 & 16 35 55.16 & +66 11 50.8 & 6.1           & 1, 2 \\
Abell 2667a & 1.035 & 23 51 40.00 & –26 04 52.0 & 17            & 1, 2 \\
The Clone   & 2.003 & 12 06 01.80 & +51 42 30.7 & 27 $\pm$ 1    & 3, 5 \\
Abell 2218a & 2.520 & 16 35 54.18 & +66 12 24.8 & 22 $\pm$ 2    & 1, 2  \\
Abell 1835a & 2.566 & 14 01 04.96 & +02 52 24.8 & 3.5 $\pm$ 0.5 & 1,
2, 6  \\
cB58       & 2.729 & 15 14 22.29 & +36 36 25.7 & $\sim30$      & 4, 7 \\
8 O'clock  & 2.731 & 00 22 40.97 & +14 31 14.0 & $\sim8$      & 8, 9
\enddata
\tablecomments{References: (1) redshifts from our LBT H$\alpha$ 
  spectroscopy; (2) \citet{Rigby08}; (3) \citet{Hainline09}; (4)
  \citet{Teplitz2000}; (5) \citet{Lin09}; (6) \citet{Smail05}; (7)
  \citet{Seitz98}; (8) \citet{Allam07}; (9) \citet{Finkelstein09}}
\label{table_sourceinfo}
\end{deluxetable*}
\end{center}

\subsection{Sample Selection}\label{sec:obs_sample}
To measure extinction and study its implication on the SFR
indicators, we need a sample of star-forming galaxies that we
understand very well. Gravitationally 
lensed galaxies are outstanding candidates because their amplified
fluxes and images enable high signal-to-noise (S/N) spectroscopy. The
candidates for our new observations were drawn from the pool of
objects studied by \citet{Rigby08}. Briefly, the \citet{Rigby08}
selection requires that objects have (1) {\it Spitzer} MIPS 24
$\micron$ flux above 0.4 mJy to be observed efficiently with IRS; (2)
morphologies in the optical {\it HST} imaging that exclude members of
the lensing cluster; (3) 
spectroscopic or probable photometric redshift above 1; (4) lensing
magnification above 3$\times$. From the \citet{Rigby08} 
sample, we further require that candidates are absent of AGN activity
based on X-ray luminosity below $10^{42}$ erg/s/cm$^{2}$ and the lack
of mid-IR power-law SEDs, indicating that their 24 $\micron$ fluxes
are not dominated by AGN emission. Apart from the objects selected
from the \citet{Rigby08} sample, we include three additional objects
that match the selection criteria; cB58, the Clone, and the 8
O'clock arc. We will discuss these objects in Section
\ref{sec:obs_lit}.

The galaxies in our sample are located behind galaxy clusters that are
well-modeled for mass distribution. The mass model of the Abell 2218
cluster \citep{Kneib96} indicates a magnification value for Abell
2218a of $22 \pm 2$ \citep{Kneib04}. For Abell 2218b and Abell 2667a,
magnification estimates of 6.1 and 17, respectively, are modeled by
\citet{Rigby08}. For Abell 1835a, we adopt the magnification estimate 
from \citet{Smail05}, $3.5 \pm 0.5$. The sample is listed in
Table \ref{table_sourceinfo}. 

\begin{figure}
\figurenum{1}
\epsscale{1.17}
\plotone{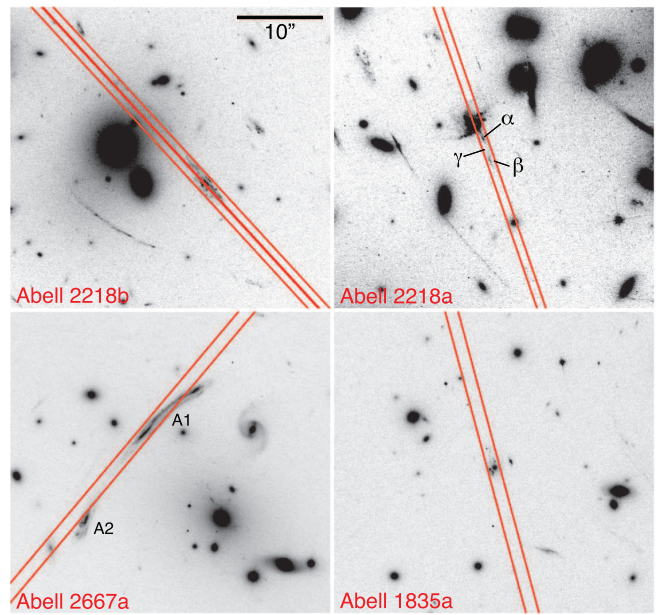}
\caption{Images of the lensed galaxies in our sample from the
  {\it Hubble} Space Telescope Advanced Camera for Survey (ACS) with
  the positions of our LBT/LUCIFER slit for NIR spectroscopy shown in
  red. Clockwise from top-left are Abell 2218b, Abell 2218a,
  Abell 1835a, and Abell 2667a. The ACS filter for Abell 2218a and
  Abell 2218b is F625W; Abell 1835 and Abell 2667 images were taken in
  F850LP. Each image is 37$''$ wide (identical scale in all four
  panels), north is up, 
  east is left. For Abell 2218a and Abell 2667a, each lensed
  component is labeled following \citet{Kneib04} and \citet{Sand05},
  respectively}
\label{fig_slitpos}
\end{figure}

\subsection{LBT/LUCIFER Near-Infrared Spectroscopy}\label{sec:obs_LBT}
Our near-IR spectroscopy was obtained in three
campaigns during 2010 October and 2011 May using the LUCIFER1
instrument on the LBT \citep{Mandel07}. The common aspects of all
these observations are that we use the N1.8 camera with plate scale of
$0\farcs25$ per pixel and with the 210 lines/mm high-resolution
grating, resulting in spectral resolution of 5.0 
\AA. A-stars were observed for spectrophotometric calibration for each 
target (details on spectral type below) at similar airmass as the
science exposure, either immediately before or after the science
exposures. Positions of the LUCIFER slit for each of our observations
are shown in Figure \ref{fig_slitpos}.

Abell 2667a was observed on 2010 November 3 in the longslit
spectroscopy mode using a slit of 1$\farcs5$ by 3.9 arcmin under
variable seeing of $1\farcs2 - 2\farcs5$. The slit was rotated 
to P.A. $= 320^{\circ}$ to place it along the length of the lensed
arc. The total integration time was 12 $\times$ 300 s with the
telescope nodded $50''$ along the slit. The unusually long nodding was
required by the $\sim 25''$ length of the arc. For calibration, we
observed HD 223466, an A3V calibration star with V $= 6.42$ mag,
through the same setup using a total integration time of $2 \times 20$
s in two nodding positions. 

Abell 2218b was observed on 2011 March 10 in the longslit spectroscopy
mode using a slit of 1$\farcs0$ by 3.9 arcmin under $0\farcs4 -
0\farcs5$ seeing. The slit was rotated to P.A. $= 42.5^{\circ}$ to
place it along the lensed arc, which is $\sim20''$ long. The arc is
$1\farcs9$ wide in the widest region; thus we took the spectra at the
central position and another position shifted by $0\farcs9$, with the
$0\farcs1$ overlapping allowing for possible pointing errors to
prevent a gap in the slit mapping. We will refer to these positions as
the ``Central'' and the ``Central+$0\farcs9$'' positions,
respectively. The total integration time was 6 $\times$ 300 s at the
Central position and 5 $\times 300$ s at the Central+$0\farcs9$
position, totalling 55 minutes, with the telescope nodded $40''$ along
the slit for both. An A0V calibration star, HD 145454, with V $= 5.43$
mag was observed using a total integration time of $2 \times 20$ s in
two nodding positions.  

Abell 2218a was observed on 2011 March 11 in the longslit spectroscopy
mode using a slit of 1$\farcs0$ by 3.9 arcmin under $0\farcs9
- 1\farcs1$ seeing. The slit was rotated to P.A. $= 18.9^{\circ}$ to
place it along the lensed arc. The total integration time was 10
$\times$ 300 s with the telescope nodded $20''$ along the slit. We
observed the same calibration star as for Abell 2218b.

Abell 1835a was observed on 2011 May 7 in the longslit spectroscopy
mode using a slit of 1$\farcs5$ by 3.9 arcmin under $1\farcs0 -
1\farcs1$ seeing. The slit was rotated to P.A. $= 15.0^{\circ}$ to
place it along the length of 
Abell 1835a. The total integration time was 12 $\times$ 
300 s with the telescope nodded $20''$ along the slit.  An
A2V calibration star, HD 122365, with V $= 5.98$ mag was observed
using a total integration time of $2 \times 20$ s in two nodding
positions.  

We took Ar lamp exposures for wavelength calibration along with dark
and flat exposures during daytime.

\begin{center}
\begin{deluxetable*}{lccccccc}
\tablewidth{0pt}
\tablecaption{Near- and Mid-Infrared Spectroscopic Observations}
\tablehead{\colhead{Source} & \multicolumn{4}{c}{IRS Exposure Time (ks)} &
  \colhead{IRS Program ID} & \colhead{Near-IR Spectroscopy} \\
 & \colhead{SL2} & \colhead{SL1} & \colhead{LL2} & \colhead{LL1} & & &}
\startdata
Abell 2218b & ... & 15.1 & 3.6\tablenotemark{a} & 3.6\tablenotemark{a} & 30775, 50586 & This Work\\
Abell 2667a & ... & 16.1 & 1.9\tablenotemark{a} & 1.9\tablenotemark{a} & 30775, 50586 & This Work \\
The Clone   & 8.6 & ... & 21.6\tablenotemark{b} & 20.9\tablenotemark{b} & 40430, 50372 & \citet{Hainline09} \\
Abell 2218a & 14.4\tablenotemark{c} & ... & ... & 3.6\tablenotemark{a} & 82, 40443 & This Work\\
Abell 1835a & 7.7 & ... & ... & 3.6\tablenotemark{a} & 82, 40443 & This Work \\
cB58       & 6.8\tablenotemark{d} & 6.8\tablenotemark{d} & 14.6\tablenotemark{c} & 34.1\tablenotemark{c} & 30832 & \citet{Teplitz2000} \\
8 O'clock & 14.4 & ... & ... & 22.1 & 40443 & \citet{Finkelstein09}
\enddata
\tablecomments{{\it Spitzer} IRS spectra previously published in (a)
  \citet{Rigby08}; (b) \citet{Fadely10}; (c) \citet{Papovich09}; (d)
  \citet{Siana08}}
\label{table_obsinfo}
\end{deluxetable*}
\end{center}

\subsection{LBT/LUCIFER Data Reduction}\label{sec:obs_LBT_reduc}
Our near-IR spectral reduction has four steps: sky subtraction,
spectral extraction, flux calibration, and line flux measurements. We
first use a modified version of an IDL longslit reduction package
written by G. D. Becker for NIRSPEC \citep{Becker09} to perform sky
subtraction following the prescription of \citet{Bian10}. The
sky-subtracted 2-dimensional spectra are corrected for the 2D
dispersion distortions, wavelength calibrated, and extracted using
IRAF. We note that the flexure compensation mechanism of LUCIFER1 was
not available during our campaigns, but clean sky subtraction was
achieved with the reduction procedure. We combine spectra for each
object by 
averaging values within each wavelength element with sigma clipping at
2.5$\sigma$ and then flux calibrate using the {\tt xtellcor\_general}
tool \citep{Vacca03} and the A-star calibrators described in the
previous section. Line fluxes for our objects are measured by
integrating the line and subtracting the background estimated 
from the mean continuum in the spectral range. We estimate the
line flux uncertainties by a Monte Carlo simulation done by repeatedly
($n = 10^4 $) simulating the object's spectrum using random values
drawn from Gaussian distribution centered at the observed flux and
$\sigma$ equal to the observed uncertainties and repeat the line
measurement. We take the resulting 1-$\sigma$ distribution of the
simulated line flux values to be the uncertainties for the line flux.

\subsection{{\it Spitzer} IRS Mid-Infrared Spectroscopy and Data
  Reduction}\label{sec:obs_IRS} 
Unpublished {\it Spitzer} IRS spectroscopy in this work is from: (1)
{\it Spitzer} program 50586 (PI G. Rieke) to take deep (integration
times $\sim$15 ks) first order Short-Low spectra (SL1; $7.4 - 14.5$
$\micron$) in the Br$\alpha$ wavelength regions of Abell 2667a and
Abell 2218b; (2) {\it Spitzer}  program 40443 (PI C. Papovich) to
take deep (integration time $7 - 14$ ks) second order Short-Low
spectra (SL2; $5.2 - 7.7$ $\micron$) in the Pa$\alpha$ wavelength
regions of Abell 2218a and Abell 1835a. For each of the four objects,
the IRS first order Long-Low spectra (LL1; $19.5 - 38.0$ $\micron$)
were from GTO programs 82 (Abell 2218a and Abell 1835a) and 30775
(Abell 2218b and Abell 2667a), PI G. Rieke. The IRS second order
Long-Low spectra (LL2; $14.0 - 21.3$ $\micron$) are available for
Abell 2218b and Abell 2667a from program 30775, and Abell 1835a from
program 82. These data are published in \citet{Rigby08}. The details
of IRS observations for our sample are summarized in Table
\ref{table_obsinfo}.   

\begin{center}
\begin{figure*}
\figurenum{2}
\includegraphics[scale=0.67,angle=90]{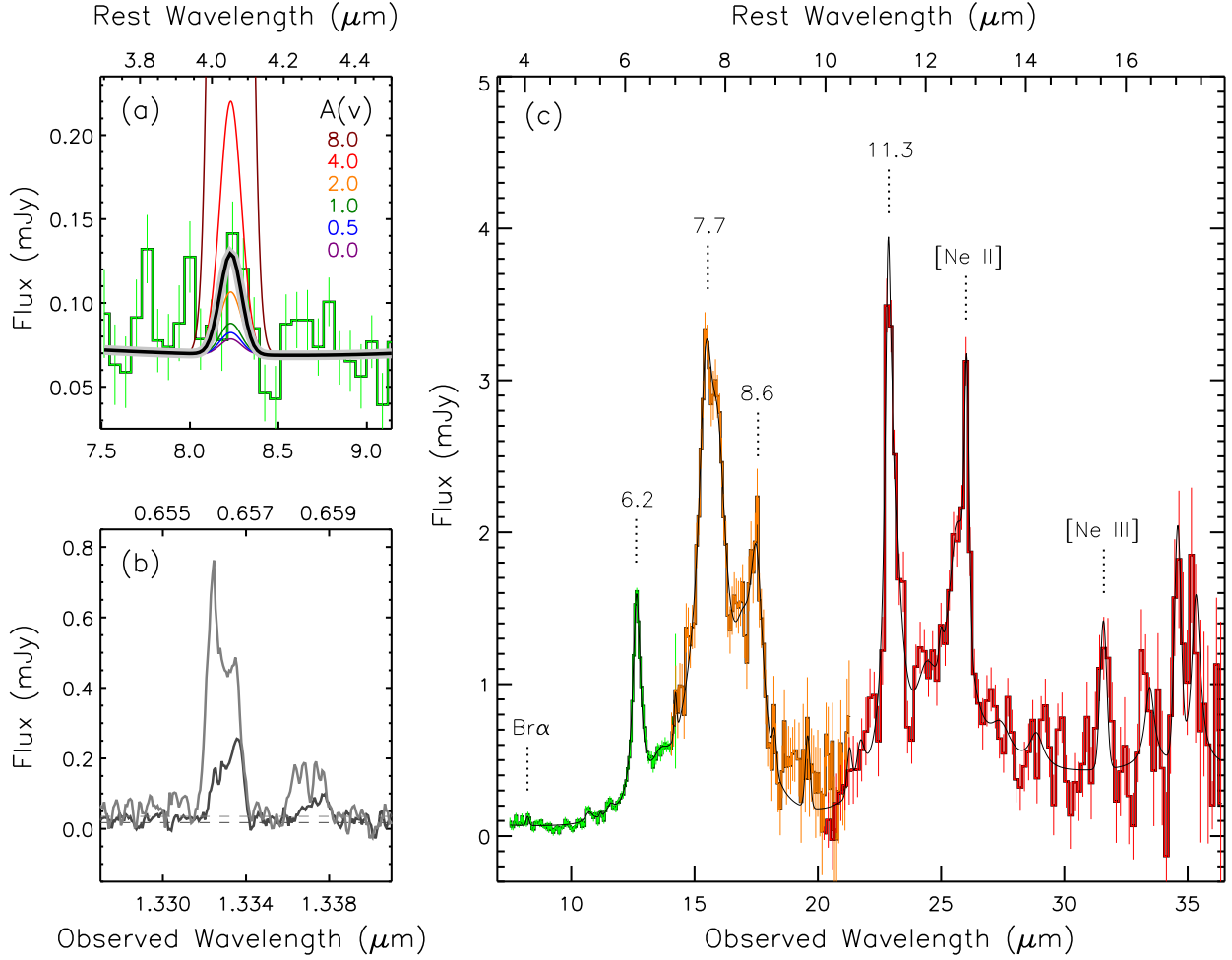}
\caption{{\it Spitzer}/IRS mid-IR and LBT/LUCIFER near-IR spectra of
  Abell 2218b. Panel (a) is the magnified region of IRS SL1 spectrum
  at the Br$\alpha$ wavelength. The best Gaussian fit to the
  Br$\alpha$ emission line is the thick black line. The color-coded
  Gaussians at the Br$\alpha$ line wavelength illustrate the expected
  Br$\alpha$ line flux for each extinction scenario from \Av\ of 0 to
  8.0 mag given the observed H$\alpha$ flux. Panel (b) shows the
  H$\alpha$ emission line from LBT/LUCIFER. The light and dark gray
  spectra are extracted from the two parallel slit positions shown in
  Figure \ref{fig_slitpos}. A double-peak line profile is clearly
  visible. The continuum level for each aperture is shown as dashed  
  lines. Panel (c) presents the overall IRS spectra with the green,
  orange, and red lines representing spectrum from the IRS SL1, LL2,
  and LL1 modules, respectively.} 
\label{fig_a2218b}
\end{figure*}
\end{center}

\begin{center}
\begin{figure*}
\figurenum{3}
\epsscale{0.85}
\plotone{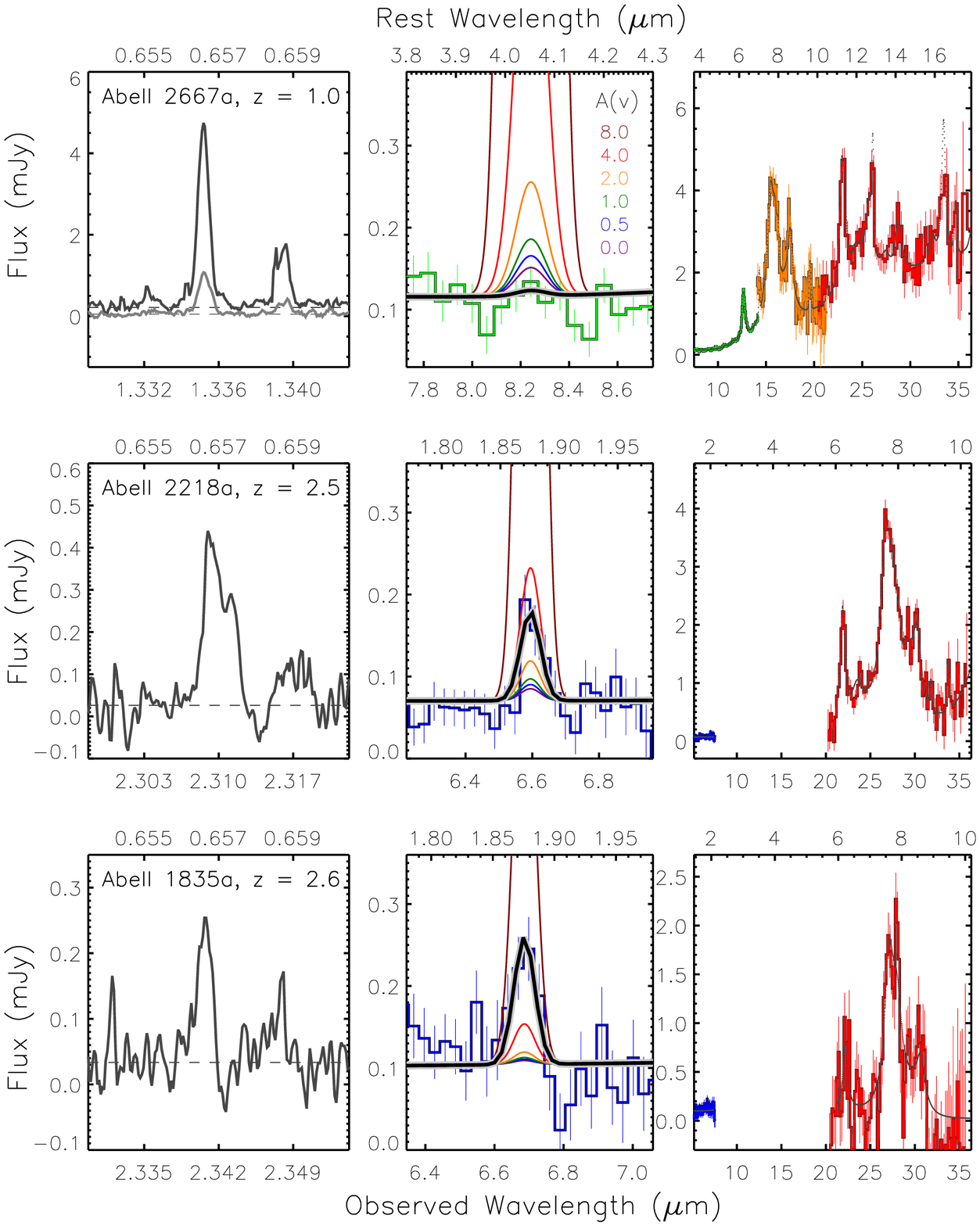}
\caption{Same as Figure \ref{fig_a2218b} but for Abell 2667a, Abell
  2218a, and Abell 1835a, from top to bottom, respectively. The IRS
  SL2 module spectra (not available for Abell 2218b in the previous
  figure), which cover Pa$\alpha$ lines for Abell 2218a and Abell
  1835a, are shown in blue. There is no SL1 and LL2 coverage for Abell
  2218a and Abell 1835a.}
\label{fig_plotall3x}
\end{figure*}
\end{center}

We have reduced all archival spectroscopic data for objects in our
sample using the most updated version of the processing software from
Level-1 data (BCD) for IRS by following the prescription of
\citet{Teplitz07}. Briefly, first we use IRAF to fit the background
slope with time for each row and subtract 
that fitted value row by row to remove latent charge. Second, we
interpolate over bad and/or hot (``rogue'') pixels using the IDL
routine {\tt IRSCLEAN} (version 2.1) provided by the SSC  based on the known
hot pixel mask for the corresponding campaign, plus manual
identification. Third, we subtract sky using a sky image constructed
from a median of other map positions. Then we co-add each map position
into a 2-dimensional (2D) spectrum and use the {\tt SPICE} software
(version 2.4) provided by the SSC to extract a 1-dimensional (1D)
spectrum using the optimal extraction template for point sources for
each map position. The 1D spectra are combined with sigma clipping at
2.5$\sigma$ into the final 1D spectra shown in Figures
\ref{fig_a2218b} and \ref{fig_plotall3x}. 

We use the PAHFIT package \citep{JDSmith07} to measure PAH feature
fluxes. PAHFIT uses $\chi^2$ minimization to simultaneously fit the
PAH and nebular emission lines, stellar and dust continua, as well as
the silicate absorption feature. We report the PAH flux measurements
in Table \ref{table_observedqn_pah}. To define the dust and stellar
continuum for the measurement of Pa$\alpha$ and/or Br$\alpha$ line
fluxes, we iteratively fit the IRS spectra with PAHFIT and interpolate 
over points with fit residuals greater than $3\sigma$. This process is
repeated until there are no remaining outliers (usually within the
3$^{\rm rd}$ iteration) and the final fit, which is effectively a
noise-free template representing the object's spectrum, is adopted as
a continuum estimate. The actual line flux measurements for both the
PAH emissions and recombination lines are done on the original
spectra, i.e. only the continua for IR recombination line
measurements are determined with this iterative fitting
procedure. This method of continuum estimation allows construction of
a mid-IR spectral template that best matches each galaxy in a
non-parametric manner. In other words, it provides continuum estimates
in the Pa$\alpha$ and Br$\alpha$ wavelength regions that are not only
constrained by photometric observations (e.g., from {\it
  Spitzer}/IRAC), but also by PAH-region information. These continuum 
estimates are shown for each object in Figures \ref{fig_a2218b},
\ref{fig_plotall3x}, and \ref{fig_litirs}.

Br$\alpha$ and Pa$\alpha$ lines are measured by fitting Gaussian
profiles with widths fixed at the resolution of the IRS module
covering the line, the central wavelengths fixed at the expected line
wavelengths based on the optical spectroscopic redshift, and the
Gaussian base given by the continuum under the expected wavelength
range of the line; the line 
peak is the only free parameter to fit. The line flux uncertainties
are estimated by a procedure similar to that of the H$\alpha$ line
flux uncertainties: (1) a series of spectra ($n = 10^4 $) were
generated randomly based on the actual spectra and the $\sigma$
uncertainties; (2) line flux measurements are carried out on them; and
(3) we take the 1-$\sigma$ value of the simulated flux distribution to
be the uncertainties for the line flux. If the formal fit to the line
yields negative line flux due to non-detection (occurring in two cases,
cB58 and The Clone), we adopt the 1-$\sigma$ value as an upper limit
to the line flux. The measurements are reported in Table
\ref{table_observedqn}.

\begin{center}
\begin{deluxetable*}{lcccccccc}
\tablewidth{0pt}
\tablecaption{Measured Fluxes}
\tablehead{\colhead{Source} & \colhead{$f$(24 $\micron$)} &
  \colhead{$f$(70 $\micron$)} & \colhead{$f$(H$\alpha$)} &
  \colhead{$f$([NII] 6583)} & \colhead{$f$(Pa$\alpha$)} &
  \colhead{$f$(Br$\alpha$)} & \colhead{$f$[Ne II 12.8]} &
  \colhead{$f$[Ne III 15.5]}\\  
& \colhead{(1)} & \colhead{(2)} & \colhead{(3)} & \colhead{(4)} &
  \colhead{(5)} & \colhead{(6)} & \colhead{(7)} & \colhead{(8)}}
\startdata
Abell 2218b & 1.67 & 7.4 $\pm$ 1.5 & 21.2 $\pm$ 0.6 & 5.2 $\pm$ 0.5 & ... & 4.3 $\pm$ 0.9 & 1.8 $\pm$ 0.2 & 0.9 $\pm$ 0.2\\

Abell 2667a & 4.52 & 19.4 $\pm$ 3.9 & 78.7 $\pm$ 1.5 & 31.1 $\pm$ 3.2 & ... & 0.4 $\pm$ 1.0 & 2.8 $\pm$ 0.4 & 0.9 $\pm$ 0.3 \\ 

The Clone   & 0.88 & 8.8 $\pm$ 1.8 & 20.2 $\pm$ 0.5\tablenotemark{a} & 3.9 $\pm$ 0.2\tablenotemark{a} & $< 1.3$ & ... & ... & ...\\

Abell 2218a & 1.16 & $<$ 7 & 6.2 $\pm$ 0.3 & 1.5 $\pm$ 0.3 & 5.7 $\pm$ 1.2 & ... & ... & ...\\

Abell 1835a & 0.99 & $<$ 13 & 1.8 $\pm$ 0.3 & 1.0 $\pm$ 0.3 & 7.9 $\pm$ 1.5 & ... & ... & ...\\ 

cB58       & 0.24\tablenotemark{b} & 1.7 $\pm$ 1.0\tablenotemark{b} & 12.6 $\pm$ 0.4\tablenotemark{c} & 1.1 $\pm$ 0.3\tablenotemark{c}  & $< 2.3$ & $< 1.3$ & ... & ...\\ 

8 O'clock  & 0.57 & ... & 17.6 $\pm$ 0.5\tablenotemark{d} & 4.8 $\pm$ 0.4\tablenotemark{d}  &
11.5 $\pm$ 2.2 & ... & ... & ...
\enddata
\tablecomments{Col. (1) 24 $\micron$ flux in mJy. $f$(24 $\micron$) of
  Abell galaxies except Abell 2667a are from \citet{Rigby08}, who
  noted that DAOPHOT errors of the flux are overly optimistic and thus
  not reported. A nominal uncertainty of 0.1 mJy is adopted for SED
  fitting for \LTIR. \citet{Rigby08} mistakingly reported a $f$(24
  $\micron$) for only one component of the A2667 arc. A correct value
  is reported here; Col. (2) 70 $\micron$ flux in mJy; Col. (3)-(4)
  optical line fluxes in 10$^{-16}$ erg/s/cm$^2$; Col. (5)-(6) IR
  recombination lines fluxes in 10$^{-16}$ erg/s/cm$^2$. The formal
  fit to the Br$\alpha$ line of cB58 yields a flux value of
  $(-0.4 \pm 1.3) \times 10^{-16}$  erg/s/cm$^2$ and hence a
  $1\sigma$ limit is reported in the Table. Also for the Pa$\alpha$
  line, the formal fit yields $(-0.2 \pm 2.3) \times 10^{-16}$
  erg/s/cm$^2$ and hence the $1\sigma$ limit is reported. Likewise, a 
  1$\sigma$ limit is reported for the Pa$\alpha$ line flux of The
  Clone, where a formal fit yields $(0.3 \pm 1.3) \times 10^{-16}$;
  Col. (7)-(8) Line fluxes of [Ne II] 12.8 $\micron$ and [Ne III] 15.5
  $\micron$ in 10$^{-15}$ erg/s/cm$^2$ measured by PAHFIT. Dots
  indicate no spectral coverage. These values are not corrected for
  lensing magnification. Flux references: (a) 
  \citet{Hainline09}; (b) \citet{Siana08}; (c) \citet{Teplitz2000};
  (d) \citet{Finkelstein09}.}
\label{table_observedqn}
\end{deluxetable*}
\end{center}

\begin{center}
\begin{deluxetable*}{lccccc}
\tablewidth{0pt}
\tablecaption{Measured PAH Fluxes}
\tablehead{\colhead{Source} & \colhead{$f$(PAH$_{6.2}$)} &
  \colhead{$f$(PAH$_{\rm 7.7~Complex}$)} & \colhead{$f$(PAH$_{8.6}$)}
  & \colhead{$f$(PAH$_{\rm 11.3~Complex}$)} & \colhead{$f$(PAH$_{12.6}$)}}
\startdata
Abell 2218b & 15.9 $\pm$ 0.2 & 72.2 $\pm$ 1.8 & 13.2 $\pm$ 1.0 & 17.3
$\pm$ 0.7 & 12.7 $\pm$ 0.9\\  

Abell 2667a & 13.0 $\pm$ 0.2 & 82.1 $\pm$ 3.0 & 19.9 $\pm$ 2.2 & 21.6
$\pm$ 1.3 & 11.7 $\pm$ 1.6\\  

The Clone   & 7.5 $\pm$ 0.3 & 15.9 $\pm$ 0.4 & 4.2 $\pm$ 0.3 & 4.5
$\pm$ 0.7 & ...\\ 

Abell 2218a & 16.0 $\pm$ 0.8 & 54.4 $\pm$ 1.3 & 10.3 $\pm$ 0.8 & ... &
... \\ 

Abell 1835a & 7.5 $\pm$ 1.4 & 26.1 $\pm$ 1.6 & 7.5 $\pm$ 1.9 & ... &
...\\  

cB58       & 2.9 $\pm$ 0.2 & 9.7 $\pm$ 0.5 & 2.5 $\pm$ 0.4 & ... &
...
\enddata
\tablecomments{PAH emission lines and emission complexes fluxes
  measured by PAHFIT in 10$^{-15}$ erg/s/cm$^2$. These values are not
  corrected for lensing magnification. Dots indicate that the feature
  is outside spectral coverage. The 7.7 $\micron$ and 11.3 $\micron$
  complexes are the sum of fluxes at wavelength $7.3-7.9$ $\micron$
  and $11.2-11.4$ $\micron$, respectively. The LL2 spectrum of the 8
  O'clock arc does not have sufficient S/N to measure PAH line fluxes
  consistently.}    
\label{table_observedqn_pah}
\end{deluxetable*}
\end{center}

\subsection{Archival and Literature Data}\label{sec:obs_lit}
We augmented our sample with three additional objects: The Clone, cB58,
and the 8 O'clock arc. These objects have suites of data similar to
our sample: rest-frame optical spectroscopy covering H$\alpha$ and
deep IRS spectroscopy covering Pa$\alpha$ (Table
\ref{table_sourceinfo}).

cB58 \citep{Yee96}, at redshift $z = 2.729$, is the first lensed LBG
known with high magnification ($\sim$30$\times$); it has been studied
extensively from UV to millimeter wavelengths (Siana et al. 2008, and
the references therein). Rest-frame optical spectroscopy was
obtained by \citet{Teplitz2000}. \citet{Siana08} present a full
suite of IRS observation (SL2, SL1, LL2, and LL1) for the galaxy. We
re-reduced these IRS data using our procedure above and found the
result to be consistent with the original reduction published in
\citet{Siana08}; details in Section \ref{sec:result_spec}. The lensing 
magnification of cB58, $\sim30\times$, was from \citet{Seitz98}.

\begin{figure}
\figurenum{4}
\epsscale{1.12}
\plotone{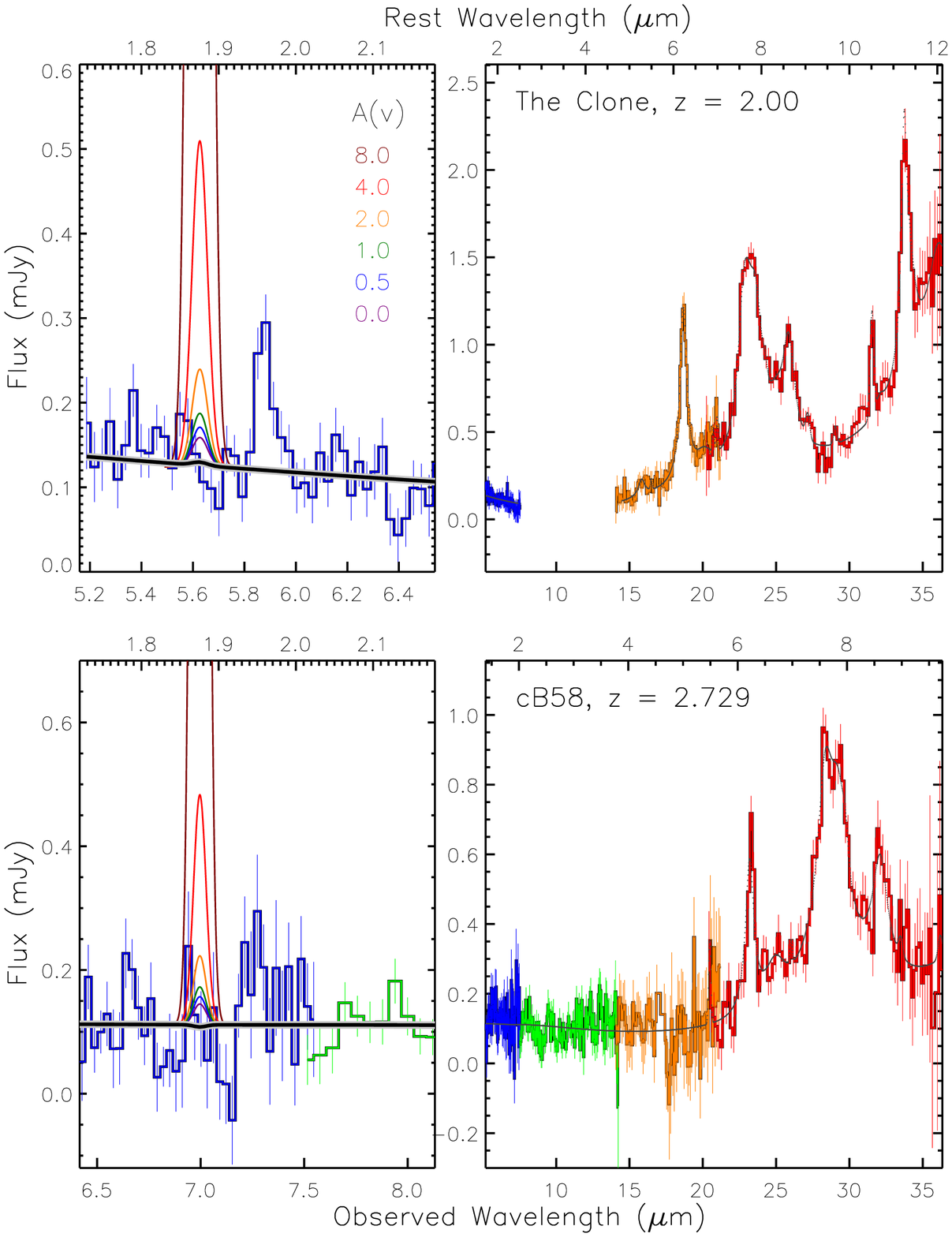}
\caption{Our reduction of the {\it Spitzer} IRS spectra for the Clone 
  (SDSSJ1206+5142) and cB58 (cf. \citet{Fadely10} and \citet{Siana08},
  respectively). The color coding for IRS modules the line fluxes for
  each \Av\ scenario is the same as Figures \ref{fig_a2218b} and
  \ref{fig_plotall3x}. We note that the apparent spectral feature at
  $\sim 5.9$ $\micron$ (observed frame) in the SL2 spectrum of the Clone
  cannot be interpreted as a Pa$\alpha$ line at the redshift $z =
  2.00$, which has been measured from multiple optical lines (see
  discussion in Section \ref{sec:result_spec}).} 
\label{fig_litirs}
\end{figure}

The Clone, at redshift $z = 2.0026$ was discovered from SDSS imaging 
\citep{Lin09}. Its rest-frame optical spectroscopy was obtained by
\citet{Hainline09}. IRS observations in the LL2 and LL1 bands were
obtained by \citet{Fadely10}, who concluded that the Clone's IR
emission is dominated by a starburst based on the strength of the 6.2
$\micron$ PAH feature, despite the object showing a very strong [SIV]
emission line. The Clone's deep IRS SL2 observation is a part of {\it
  Spitzer} program 50372 (PI C. Papovich) with an integration time of
8.6 ks. As with cB58, we re-reduced all IRS data using our procedure
and software to ensure consistency. We present our reduction of these
archival data in Figure \ref{fig_litirs}. The lensing magnification
for the Clone, $27 \pm 1 \times$, was reported by \citet{Lin09}.

The 8 O'clock arc, at redshift $z = 2.7308$ was discovered
serendipitously from SDSS imaging by
\citet{Allam07}. \citet{Finkelstein09} obtained the rest-frame
spectroscopy of components `A2' and `A3' in the nomenclature of
\citet{Allam07}. The arc was observed as a part of {\it Spitzer}
program 40443 (PI C. Papovich) with an SL2 integration time of 14.4
ks. The slit for the SL2 observation is centered on the `A2' component
and  
covered significant parts of both the `A1' and `A3' components such
that any bias due to spatial variation of recombination lines along
the lensed arc will be negligible. The PAH spectrum of the 8 O'clock
arc shows 6.2 $\micron$ and 7.7 $\micron$ emission features, yet at
low S/N due to contamination in the IRS slit from a nearby IR-bright
dust-obscured galaxy, which we will discuss in the next section. We
adopt the magnification estimate of $8\times$ for the arc
\citep{Finkelstein09}. 

\subsection{Additional {\it Spitzer} MIPS Photometry and Data
  Reduction}\label{sec:obs_MIPS} 
The {\it Spitzer} MIPS 24 and 70 $\micron$ photometry for objects 
behind the Abell clusters are taken from \citet{Rigby08} with
exception of Abell 2667a, whose 70 $\micron$ photometry is reported as
an upper limit in \citet{Rigby08} but which was re-observed with
longer integration  time in {\it Spitzer} program ID 50586 (PI
G. Rieke). For the Clone, we downloaded the 24 and 70 $\micron$ data
observed as part of programs 40430 (PI S. Allam) and 50372 (PI
C. Papovich). The {\it Spitzer} MIPS observations for the 8 O'clock
arc were obtained under program 40443 (PI C. Papovich).

We measured the 24 and 70 $\micron$ fluxes from the Level-2 (PBCD)
data using the {\tt Starfinder} IDL routine \citep{Diolaiti2000} to
perform PSF photometry. The PSF model was generated using the {\tt
  STinyTim} routine provided by the SSC and smoothed according to a
prescription given by \citet{Engelbracht07} to better match the
observed PSF. We require $\sigma = 3$ and the PSF correlation value of
0.75 for our 24 and 70 $\micron$ flux measurements. The 8 O'clock arc
has a 24 $\micron$-bright dust-obscured galaxy $\sim6\farcs5$ away
that partially blends with the arc, but we were able to deblend the
flux by performing PSF photometry to subtract the intervening source
first and repeat the photometry to measure the flux of the arc. This
contaminating galaxy is highly extincted, such that it is undetected
in the SDSS imaging and was discovered only in the 24 $\micron$
observation. Our flux measurements along with fluxes from the
literature are reported in Table \ref{table_observedqn}.

\begin{figure}
\figurenum{5}
\epsscale{1.1}
\plotone{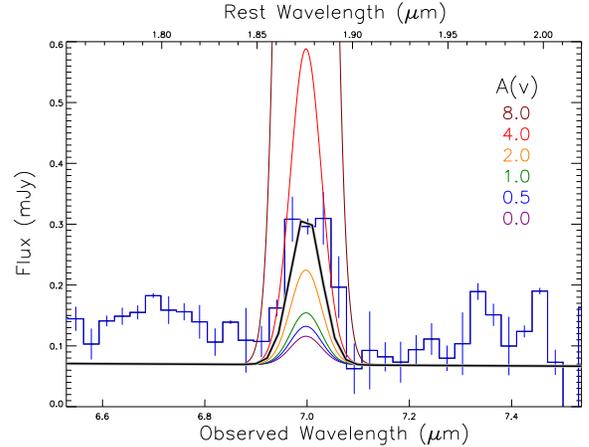}
\caption{The {\it Spitzer} IRS SL2 spectrum for the 8 O'clock arc,
  showing the Pa$\alpha$ line. The color coding for IRS module and the
  line fluxes for each extinction scenario are the same as Figures
  \ref{fig_a2218b}, \ref{fig_plotall3x}, and \ref{fig_litirs}.}  
\label{fig_8oclock}
\end{figure}

\subsection{Augmented Sample Summary}\label{sec:obs_sum}
Our augmented sample consists of the four galaxies we observed plus
the three from the literature. In addition to the measurements of
H$\alpha$ and of at least one of the IR recombination lines
(Pa$\alpha$ and/or Br$\alpha$), all of them have good 24 $\micron$
photometry, far-IR and sub-mm photometry as well as IRS spectroscopy
covering, at least, the $6 - 8$ $\micron$ wavelength region. The IRS
coverage goes out to nearly 17 $\micron$ (rest-frame) in the two
lowest redshift objects (Abell 2667a and Abell 2218b, $z \sim 1.0$).

\begin{center}
\begin{deluxetable*}{lccccccccccccc}
\tablewidth{0pt}
\tablecaption{Derived Quantities}
\tablehead{Source & \multicolumn{2}{c}{LIR (log\Lsun)} &
  \multicolumn{5}{c}{SFR (\Msun/yr)}
  & A$_{\rm v}^{{\rm H}\alpha/{\it Spitzer}}$ & A$_{\rm v}^{{\rm IR}}$ &
  A$_{\rm v}$\\  
    & 24 $\micron$ & Far-IR SED &   H$\alpha$ & H$\alpha$ + 24
  $\micron$ & PAH$_{6.2~\micron}$ & LIR$_{24~\micron}$ & LIR$_{{\rm
      Far-IR SED}}$ & & & & \\
& (1) & (2) & (3) & (4) & (5) & (6) & (7) & (8) & (9) & (10)} 
\startdata
Abell 2218b & 11.69 & 11.68 & 10.3 & 36.6 & 66.7 & 48.0 & 46.8 & 2.8 $\pm$ 0.3 & 2.1 & 2.5\\

Abell 2667a\tablenotemark{$\star$} & 11.68 & ...   & 13.9 & 39.9 & 19.4 & 47.1 & ... & -2.4\tablenotemark{$\star$} $\pm$ 3.1 & 1.7 & -0.4\\ 

The Clone\tablenotemark{$\dagger$}   & 11.20 & 11.38 & 11.4 & 20.0 & 36.2 & 15.1 & 23.2 & $< -1.1$\tablenotemark{$\dagger$} & 1.0 & 1.0\\ 

Abell 2218a & 11.81 & 11.83 & 7.5  & 42.3 & 170.2 & 63.8 & 66.5 & 3.3 $\pm$ 0.4 & 3.0 & 3.2 \\

Abell 1835a & 12.71 & 12.83 & 14.5 & 298 & 558.2 & 561 & 752 & 5.9 $\pm$ 0.4 & 5.4 & 5.7\\

cB58\tablenotemark{$\dagger$} & 11.04 & 11.04 & 13.6 & 19.8 & 26.6 & 10.4 & 10.4 & $< 0.7$\tablenotemark{$\dagger$} & -0.4 & -0.4\\ 

8 O'clock & 12.17 & ... & 71.5 & 154 & ... & 149 & ... & 2.8 $\pm$ 0.3 & 1.0
& 1.9
\enddata
\tablenotetext{$\star$}{IR recombination line detected at low significance.}
\tablenotetext{$\dagger$}{IR recombination line not detected.}
\tablecomments{Quantities in this table are corrected for lensing
  magnification. Col. (1) \LTIR\ estimated from the monochromatic 24
  $\micron$ indicator from \citet{Rujopakarn12}; Col. (2)
  \LTIR\ measured by fitting SED to far-IR and sub-mm photometry. The
  values for Abell 2218b and Abell 1835a are from \citet{Rigby08}. The
  values for Abell 2218a, the Clone and cB58 are from
  \citet{Finkelstein11}, \citet{Fadely10} and \citet{Siana08},
  respectively.; Col. (3) H$\alpha$ SFR estimated by the
  \citet{Kennicutt98} formula (no extinction correction); Col. (4)
  Extinction-corrected SFR estimated using the H$\alpha$ and
  rest-frame 24 $\micron$ luminosity formula provided by
  \citet{Kennicutt09}; Col. (5) Col. SFR estimated by the PAH 6.2
  $\micron$ complex luminosity; (6) SFR estimated by 24 $\micron$
  monochromatic indicator provided by \citet{Rujopakarn12}; Col. (7)
  SFR estimated using the Far-IR \LTIR\ in Col. (2) and the
  relationship given by \citet{Rujopakarn12}; Col. (8) Screen
  extinction \Av\ in mag estimated with H$\alpha$ and IR recombination
  lines assuming case-B  recombination scenario. The value for cB58
  uses H$\alpha$ and the Pa$\alpha$ line, the Br$\alpha$ limit for
  cB58 indicates an extinction limit of \Av\ $< 1.8$ mag.; Col. (9)
  \AvIR\ in  mag. See Section \ref{sec:result_ext} for more details.;
  Col. (10) Final extinction estimate from averaging values in Col (8)
  and (9). For cB 58 and The Clone, the \AvIR\ values are adopted.}  
\label{table_derivedqn}
\end{deluxetable*}
\end{center}

\section{RESULTS}\label{sec:result}
In this section, we describe the near- and mid-IR spectra and the
derived physical quantities, including metallicity, \LTIR\ and SFR
from optical and IR tracers, and extinction measurements.

\subsection{Near- and Mid-IR Spectroscopy}\label{sec:result_spec}
The final 1D H$\alpha$ spectra from our LBT/LUCIFER1 observations are
shown in Figures \ref{fig_a2218b} and \ref{fig_plotall3x} along with
the IRS spectra. The 6.2 $\micron$ PAH emission line, 7.7 $\micron$
complex, and 8.6 $\micron$ complex are clearly detected in all
galaxies.

The spectra from both slit positions (Section \ref{sec:obs_LBT}) of
Abell 2218b show highly asymmetric line profiles. The profile at the
central position is double-peaked in both H$\alpha$ and [NII] while
the central+$0\farcs9$ spectrum shows a single-sided profile with
stronger emission at the red side of the line. The combined line
profile from both slit positions is well fitted with a double
Gaussian, whose deconvolved spectral width (corrected for instrumental
resolution) 
indicates a velocity dispersion of $163 \pm 3$ km/s. Traces of the arc
northeast of  the main component hinted in the ACS image are not
detected at H$\alpha$. The IRS spectrum for Abell 2218b shows a strong
[Ne III] 15.56 $\micron$ line as well as the [Ne II] line at 12.81
$\micron$, apart from other PAH features. The Br$\alpha$ line is
detected at $4.8\sigma$ with line flux of 4.3 $\pm$ 0.9 $\times$
10$^{-16}$ erg/s/cm$^2$. 

The spectra of Abell 2667a were extracted from two apertures in the
slit. Both of these positions are very bright in H$\alpha$. Referring
to Figure \ref{fig_slitpos}, these apertures are at (1) the main arc
north-northeastern of the brightest cluster galaxy (BCG) (identified
as `A1' by \citet{Sand05}, and in Figure \ref{fig_slitpos}); and (2)
the smaller arc east of the  
BCG (the `A2' arc in the Sand et al. study). Although it appears that
a part of the smaller arc is outside the slit in the optical image,
the H$\alpha$ line is clearly detected (the light grey spectrum in
Figure \ref{fig_plotall3x} top row, left column). Both apertures show
the same profile for both H$\alpha$ and [NII] lines. No velocity shift
between the two components is observed. The formal fit to the
Br$\alpha$ line yields a flux of $(0.4 \pm 1.0) \times 10^{-16}$, a
non-detection. The galaxy has the warmest dust continuum among our
samples, as seen in the rising SED towards the red end of the IRS
spectrum in Figure \ref{fig_plotall3x} (cf. Abell 2218b's continuum in
Figure \ref{fig_a2218b}). However, the strong PAH emission is
inconsistent with the system's emission being dominated by an AGN in
the mid-IR. Its log([NII]/H$\alpha$) line ratio of $-0.40$ alone
(i.e. without the [OIII] and H$\beta$) does not unambiguously support
or rule out the presence of a weak narrow-line AGN. But even with the
presence of a weak, optically selected AGN, the EW of the 6.2 micron
PAH emission could still serve as a good tracer of SFR because the
suppression of PAH emission due to AGN is found to be limited to the
nuclear region \citep{DiamondStanic10}.

The H$\alpha$ line of Abell 2218a also shows a double-peaked
profile indicating a velocity dispersion of $188 \pm 5$ km/s. The
galaxy has been a subject of a previous near-IR spectroscopic study by
\citet{Richard11}. Our slit orientation, however, is different (see
their Figure 1); they excluded the $\beta$ and 
$\gamma$ components (the former being a major component) in the
nomenclature of \citet{Kneib04}, shown in Figure \ref{fig_slitpos},
while our slit orientation covers all the main components ($\alpha$,
$\beta$ and $\gamma$). The Pa$\alpha$ observation, which targets the
$\beta$ component, detected Pa$\alpha$ line at $4.8\sigma$ with line
flux of 5.7 $\pm$ 1.2 $\times$ 10$^{-16}$ erg/s/cm$^2$. This
measurement is consistent with the previously published reduction of
the same data by \citet{Papovich09} who measured the line flux to be
8.5 $\pm$ 1.4 $\times$ 10$^{-16}$ erg/s/cm$^2$. We attribute the small
difference in the two measurements to the different definition of the
continuum: \citet{Papovich09} defined their continuum with IRAC
photometry whereas our reduction adopted the simultaneous
multi-component fit of spectral features described in Section
\ref{sec:obs_IRS}.

Abell 1835a shows a faint H$\alpha$ line, too faint to measure 
asymmetry in the line profile, but a bright Pa$\alpha$ line with a
flux of 7.9 $\pm$ 1.5 $\times$ 10$^{-16}$ erg/s/cm$^2$. This is the 
second brightest Pa$\alpha$ detection in our sample, after the 8
O'clock arc, whose Pa$\alpha$ flux is 11.5 $\pm$ 2.2 $\times$
10$^{-16}$ erg/s/cm$^2$. The LL1 PAH spectrum of the 8 O'clock arc
does not have sufficient S/N to use PAHFIT to construct a noise free 
spectrum as a template for the Pa$\alpha$ line flux measurement
(Section \ref{sec:obs_IRS}) due to the contamination from the nearby
IR-bright dust-obscured galaxy, thus instead the continuum of the arc 
for the measurement is defined by interpolating between the IRAC 5.0
and 8.0 $\micron$ photometric points. These IRAC observations yield
fluxes of $74 \pm 3$ and $64 \pm 4$ $\mu$Jy for the 5.8 and 8.0
$\micron$ bands, respectively. We present the Pa$\alpha$ spectrum of
the 8 O'clock arc in Figure \ref{fig_8oclock}.

For the two IRS spectra from the literature, the Clone and cB58, our
reduction generally agrees with those published previously by
\citet{Fadely10} and \citet{Siana08}, respectively. Small differences
are noted in the PAH line fluxes due to the way that the continuum for
the fit is defined. PAHFIT uses multi-component dust features to
represent the continuum, while a single-component power-law is used by
Fadely et al. and Siana et al. Their approach resulted in a similarly
good fit to the data compared to ours, although the measured PAH flux
could differ, as observed in the Clone's PAH fluxes from the
$6.2-11.3$ $\micron$ features where the \citet{Fadely10} values are a
factor of $2.6-3.7$ larger than ours. Fadely et al. reported that
their results are a factor of $1-8$ higher than those of
\citet{Brandl06} in their re-reduction of the same data set. Brandl et
al. use a continuum level defined on either side of the emission
features and thus their PAH fluxes will be systematically smaller than
with either our PAHFIT measurement or the single component continuum
method. That is, we expect the measurements of the PAH line flux from
the same data to yield flux values in the following increasing order:
Brandl et al., PAHFIT, and Fadely et al./Siana et al., which is
consistent with the differences we find. The formal Pa$\alpha$ line
fit for cB58 yields a flux of $(-0.2 \pm 2.3) \times 10^{-16}$
erg/s/cm$^2$, from which we determined the Pa$\alpha$ flux limit to be
$< 2.3 \times$ 10$^{-16}$ erg/s/cm$^2$, which agrees with the
non-detection reported by \citet{Siana08}, $< 6 \times 10^{-16}$
erg/s/cm$^2$. The formal Br$\alpha$ line fit yields a line flux of
$(-0.4 \pm 1.3) \times 10^{-16}$ erg/s/cm$^2$, thus we report a limit
of $< 1.3\times 10^{-16}$ erg/s/cm$^2$.

The formal fit to Pa$\alpha$ line flux of the Clone yields $(0.3 \pm
1.3) \times 10^{-16}$ erg/s/cm$^2$, from which we report an upper limit
similar to that of cB58 above. In the panel showing the Pa$\alpha$
wavelength region for the Clone in Figure \ref{fig_litirs}, we note a
spectral feature at 5.9 $\micron$ (1.96 $\micron$ rest-frame) that
resembles a Pa$\alpha$ line at $z \sim 2.1$. However, the
spectroscopic redshift of 2.0026 was measured with multiple optical
lines \citep{Hainline09} and thus precludes the possibility of the
feature being interpreted as the Pa$\alpha$ line. It is also unlikely
to be an H$_2$ emission line at 1.96 $\micron$ given the absence of
other H$_2$ lines nearby (2.03 and 2.12 $\micron$). The source of the
apparent feature therefore remains unknown.

\subsection{Metallicity}\label{sec:result_metallicity}
We estimate metallicity with the N2 index \citep{PettiniPagel04}
using both linear and 3$^{\rm rd}$-order fits. The index is not
affected by the optical extinction given the proximity of the
lines. Both estimators yield consistent values, which agree well with
those found by \citet{Erb06} for massive galaxies at $z \gtrsim 2$
selected via UV luminosity. The range of the metallicity found in our
sample is also in good agreement with the sample of 28 lensed galaxies
at $1.5 < z < 5$ observed by \citet{Richard11}, $8.00 \geq Z \geq
8.94$, with an average oxygen abundance of $Z = 8.55$, similar to the
average value from the linear N2 index in this work, $Z = 8.56$. That
is, our sample has typical oxygen abundance for field and lensed
galaxies at similar $z$.

\begin{center}
\begin{deluxetable}{lcc}
\tablewidth{0pt}
\tablecaption{Metallicity Estimates from [NII]/H$\alpha$}
\tablehead{Source & Z$_{\rm PP04}$ & Z$_{\rm PP04,~3^{\rm rd}~order}$\\  
& (1) & (2)} 
\startdata
Abell 2218b & 8.55 $\pm$ 0.04 & 8.53 $\pm$ 0.10\\
Abell 2667a & 8.67 $\pm$ 0.05 & 8.73 $\pm$ 0.10\\
The Clone   & 8.49 $\pm$ 0.02 & 8.45 $\pm$ 0.07\\
Abell 2218a & 8.55 $\pm$ 0.10 & 8.52 $\pm$ 0.25\\ 
Abell 1835a & 8.75 $\pm$ 0.15 & 8.91 $\pm$ 0.33\\
cB58        & 8.31 $\pm$ 0.10 & 8.26 $\pm$ 0.35\\
8 O'clock   & 8.58 $\pm$ 0.04 & 8.57 $\pm$ 0.09
\enddata
\tablecomments{Values are in $12 +$ log(O/H). Col. (1) Metallicity
  estimates from the N2 linear formula provided by
  \citet{PettiniPagel04}; Col. (2) Metallicity estimates from the N2
  3$^{\rm rd}$-order estimator from \citet{PettiniPagel04}.} 
\label{table_metal}
\end{deluxetable}
\end{center}

\subsection{Infrared Luminosity and Star Formation
  Rate}\label{sec:result_lir} 
We can measure the star-formation rate (SFR) from \LTIR, which in
turn can be estimated by three tracers: (1) far-IR SED fitting; (2)
the monochromatic 24 $\micron$ \LTIR\ estimator given 
by \citet{Rujopakarn12}; (3) PAH emission line luminosity. The SFR
can also be estimated from the H$\alpha$ luminosity as well as the
combination of H$\alpha$ and 24 $\micron$ luminosity. We adopt the SFR
derived from the SED-fitted \LTIR\ as a fiducial for
comparison. SFRs are calculated from \LTIR\ by the relationship given
by \citet{Rujopakarn12}. All SFR estimates are converted from their
respective IMF assumptions to the \citet{Kroupa02} IMF. The SFR
results are presented in this section and their implications are
discussed in Section \ref{sec:discuss}.

The far-IR SED fitting (e.g., 30 $\micron$ to $\sim$1 mm) is the most
straightforward among the methods to measure \LTIR. However,
far-IR photometry is often affected by confusion noise at longer
wavelengths in crowded fields, thus limiting the applicability of
far-IR SED fitting in deep cosmological surveys. The far-IR and 
sub-mm photometry in our sample is aided by magnification of the
gravitational lenses and thus we can probe fluxes below the typical  
confusion limit of the instrument. The SED fitting for the \LTIR\ for
Abell 2218b and Abell 1835a were carried out by \citet{Rigby08} by
combining the 24 and 70 $\micron$ data with submillimeter
observations, including at least 450 and 850 $\micron$ for all
their galaxies and additionally 1.3 mm for Abell 1835a. The value for
Abell 2218a is measured using {\it Herschel} observations by
\citet{Finkelstein11}. The SED-fitted \LTIR\ for the Clone and cB58
were from the \citet{Fadely10} and \citet{Siana08} 
studies, respectively. Abell 2667a and the 8 O'clock arc have no
\LTIR\ estimates from far-IR SED fitting because the longest
band we presently have is 70 $\micron$. The 70 $\micron$ imaging for
the latter is complicated the a nearby bright dust-obscured galaxy
(Section \ref{sec:obs_MIPS}) that blends with the arc.

The monochromatic (i.e. single-band) 24 $\micron$ \LTIR\ estimator
from \citet{Rujopakarn12} yields \LTIR\ consistent with the far-IR
SED fitting at $0 < z < 2.8$ within 0.02 dex, on average, with a
scatter of  0.12 dex for individual star-forming
galaxies. \citet{Rujopakarn12} take into account the evolution of the
bolometric correction from local galaxies out to $z = 2.8$ (the
farthest redshift where the 24 $\micron$ band traces predominantly
dust and PAH emissions) due to the structural differences of IR
galaxies by using the IR surface brightness, \LIRSD, as an indicator
of IR SED, which has been demonstrated by
\citet{Rujopakarn11}. Specifically, high-$z$ IR galaxies are
physically extended and typically have $100 - 1000\times$ more surface
area than local counterparts at similar \LTIR\ \citep{Rujopakarn11},
and thus larger PAH-emitting area, which affects the bolometric
corrections in the PAH wavelength region that is probed by the 24
$\micron$ band, particularly at $z > 1$ \citep{Rujopakarn12}. We
report the values of \LTIR\ from this 24 $\micron$ indicator along
with those from SED fitting in Table \ref{table_derivedqn}. The
agreement between the monochromatic indicator and the far-IR SED
fitting is within 0.12 dex (average difference of 0.06 dex), similar
to the agreement reported in \citet{Rujopakarn12}. No systematic
trends in redshift, luminosity, or metallicity are observed.

\begin{figure}
\figurenum{6}
\epsscale{1.15}
\plotone{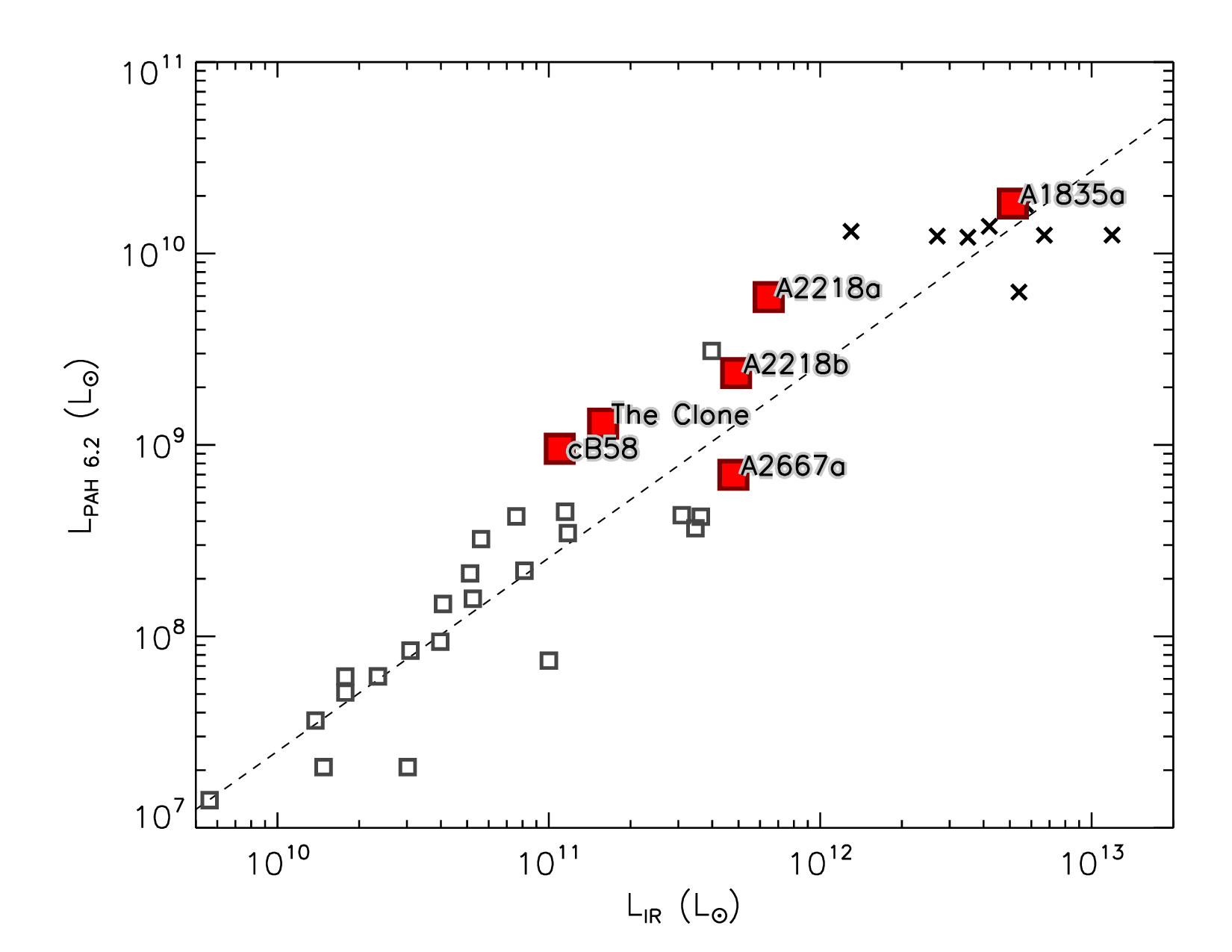}
\caption{The relationship of 6.2 $\micron$ PAH emission line
  luminosity to \LTIR\ in our sample (large red squares) compared to
  the local star-forming galaxies and LIRGs from the \citet{Brandl06}
  study (squares) and the high-redshift sub-mm galaxies from
  \citet{Pope08} (crosses). The dashed line is a linear relationship
  between log luminosities with slope of 1.01 measured by
  \citet{Pope08}. Our galaxies fall within this relationship,
  confirming the approximately unity slope reported by Pope et al. at
  the intermediate luminosities between the local and SMG samples.}
\label{fig_Pope}
\end{figure}

The last \LTIR\ estimator is derived via the luminosity of the 6.2
$\micron$ PAH emission line, following the 
relationship given by \citet{Pope08}, which was measured by combining
a local sample from \citet{Brandl06} and the results for their
sub-millimeter galaxies (SMGs). \citet{Pope08} report a slope
of the relationship between log 6.2 $\micron$ PAH luminosity and log
\LTIR\ to be approximately unity (formal slope = $1.01 \pm 0.01$) over
\LTIR\ ranging from $10^{10}-10^{13}$ \Lsun. Our objects are shown in
Figure \ref{fig_Pope} on top of data points from \citet{Pope08} and
\citet{Brandl06}. The 24 $\micron$ \LTIR\ values are used here instead
of those from the far-IR SED fitting because Abell 2667a has no far-IR
\LTIR\ estimate. Because of the methodological differences in the PAH
flux measurements between \citet{Pope08} and ours, particularly the
assumption of the continuum level, for the purpose of the comparison
in Figure \ref{fig_Pope}, we normalize the PAH luminosity by a factor
determined by matching our 6.2 $\micron$ PAH luminosity of 
cB58 to the measurement of the same object by \citet{Siana08}, who
adopt the same continuum assumption as \citet{Pope08}. The 6.2
$\micron$ PAH feature was chosen to represent the aromatic emissions
over other stronger PAH emission lines and emission complexes because
it is narrow and relatively isolated from other lines, which mitigates 
the systematic uncertainties introduced by the process of deblending
contributions from adjacent lines (which is required for, e.g., the
7.7 $\micron$ complex and 8.6 $\micron$ line). To estimate the SFR
from the 6.2 $\micron$ PAH luminosity, the PAH luminosity is first
converted to \LTIR\ via the \citet{Pope08} conversion and then to SFR
via the \citet{Rujopakarn12} relation.

In addition to estimating SFR via \LTIR, we also show: the
\citet{Kennicutt98} formula for H$\alpha$ (no extinction correction)
and the \citet{Kennicutt09} indicator that combines H$\alpha$ and the
rest-frame 24 $\micron$ luminosity in order to correct for dust
extinction in the SFR estimates. The SFR estimates from each method
are tabulated in Table \ref{table_derivedqn}. We discuss the effects
of extinction on the SFR estimates in Section \ref{sec:discuss}. 

\subsection{Extinction}\label{sec:result_ext}
Extinctions derived from optical lines (e.g. H$\alpha$/H$\beta$) could
suffer systematic underestimation where the overall extinction is
large and the dust is mixed with the sources. This possibility can be
tested with IR recombination lines which are more robust to high
extinction \citep{AH06}. To obtain a fiducial estimate, we assume a
foreground screen and measure the extinction relative to H$\alpha$
using the Pa$\alpha$ and Br$\alpha$ lines assuming case-B
recombination. For instance, for Br$\alpha$, the A$_{\rm v}^{{\rm
    H}\alpha/{\it Spitzer}}$ is given by  
\begin{equation}
\label{eq1}
\frac{I_{{\rm Br}\alpha}}{I_{{\rm H}\alpha}} =
\frac{I_{{\rm Br}\alpha,0}}{I_{{\rm H}\alpha,0}}~{\rm exp}\left[
  -\frac{A_{\rm v}^{{\rm H}\alpha/{\it Spitzer}}}{1.086}(A_{{\rm Br}\alpha}
  - A_{{\rm H}\alpha})\right] 
\end{equation}

\noindent where $I_{{\rm Br}\alpha,0} / I_{{\rm H}\alpha,0} = 0.0291$
and for Pa$\alpha$, $I_{{\rm Pa}\alpha,0} / I_{{\rm H}\alpha,0} =
0.1226$ \citep{Osterbrock} for T$_e =$ 10,000 K at the low-density
limit. The A$_{{\rm H}\alpha}$, A$_{{\rm Pa}\alpha}$, and A$_{{\rm
    Br}\alpha}$ are given by interpolation using the
\citet{RiekeLebof85} extinction law, which for \Av\ of 1 mag are 0.8
mag, 0.15 mag, and 0.04 mag, respectively.

Alternatively, the extinction can be estimated from the optical and IR 
SFR values by taking the latter to be a fiducial SFR. The assumption is
secure in our sample of IR luminous galaxies, where direct UV leakage 
from the galaxy is expected to be small (Rieke et al. 2009 and
references therein). This alternative 
extinction estimate, \AvIR, following \citet{Choi06} is
\begin{equation}
\label{eq2}
{\rm A}_{{\rm H}\alpha}^{\rm IR} = 2.5~{\rm log}\left(\frac{{\rm
    SFR}_{{\rm IR}}}{{\rm SFR}_{{\rm H}\alpha}} \right)
\end{equation}

\noindent The A$_{{\rm H}\alpha}^{\rm IR}$ can then be converted to
the \AvIR\ via the \citet{RiekeLebof85} extinction law. When
available, we use the far-IR SFR$_{\rm IR}$ to calculate the \AvIR,
otherwise the value estimated from the 24 $\micron$ \LTIR\ is used.

The A$_{\rm v}^{{\rm H}\alpha/{\it Spitzer}}$ method assumes a
foreground screen of dust and therefore is a lower limit, whereas the
A$_{\rm v}^{{\rm IR}}$ does not. The comparison of measurements from
both methods, tabulated in Table \ref{table_derivedqn}, shows that in 
three out of four galaxies where IR recombination lines are well
detected (Abell 2218b, Abell 2218a, and Abell 1835a), both methods agree within the range of uncertainty, which suggests that the nature of the dust 
distribution in these galaxies is roughly uniform, resembling the the
foreground screen assumption. The other object with a well-detected
Pa$\alpha$ (the 8 O'clock arc), however, shows a 1.8 mag difference
between A$_{\rm v}^{{\rm H}\alpha/{\it Spitzer}}$ and A$_{\rm v}^{{\rm
    IR}}$. The latter disagreement indicates an inhomogeneous mixture
of dust in the 8 O'clock arc and highlights the diversity of the dust
distribution scenarios at redshift $1 < z < 3$.

Abell 2218a, Abell 1835a, cB58, the Clone, and the 8 O'clock arc have
extinction measurements from optical and/or rest-frame optical
spectroscopy in the literature that can be compared with our IR 
measurements. \citet{Richard11} found $E(B-V)_{\rm star}$ of 0.18 for
Abell 2218a from SED fitting, implying \Av\ of 0.6 mag, assuming the
$R = 3.1$ law. Abell 1835a has extinction measurements by
\citet{Nesvadba07} using H$\alpha$/H$\beta$ ratio of $E(B-V) =
1.3-1.6$, implying \Av\ of $4.0-5.0$ mag. \citet{Teplitz2000} measured
$E(B-V)$ of 0.27 for cB58 from the H$\alpha$/H$\beta$ line ratio,
implying \Av\ of 0.4 mag. For the Clone, \citet{Hainline09} measured
$E(B-V)$ using the H$\alpha$/H$\gamma$ line ratio to be 0.28, which
implies \Av\ of 0.9 mag. The extinction of the 8 O'clock arc were
measured by \citet{Finkelstein09} using weighted mean of to H$\alpha$,
H$\beta$, and H$\gamma$ line ratios be $\Av\ = 1.17 \pm 0.36$ mag
(using only H$\alpha$/H$\beta$ yields $E(B-V)_{\rm gas}$ of 0.97,
implying \Av\ of 1.3 mag); and by \citet{DZ11} to be $E(B-V)_{\rm
  gas} = 0.30 \pm 0.04$ mag, implying \Av\ of 0.9 mag. From Table
\ref{table_derivedqn}, the extinction measurements from IR
recombination lines for Abell 2218a, Abell 1835a, and the 8 O'clock
arc, where we have secure IR line detections are $3.3 \pm 0.4$ mag,
$5.9 \pm 0.4$ mag, and $2.8 \pm 0.3$ mag, respectively. While our
sample size is too small to draw a general conclusion, the
comparison suggests that the optical measurements may sample
systematically lower extinction regions, in agreement with local LIRGs
and ULIRGs \citep[e.g.,][]{AH06}. 
 
The A$_{\rm v}^{{\rm H}\alpha/{\it Spitzer}}$ values for Abell 2667a
and the Clone are negative because the formal fits to the Pa$\alpha$
and Br$\alpha$ line yield values (or flux limits) lower than those
expected for the extinction of 0 mag (as shown in the simulated
\Av\ in Figures \ref{fig_plotall3x} and \ref{fig_litirs}); the line is
undetected), albeit with a considerable uncertainty. For cB58, the
line flux for both Pa$\alpha$ and Br$\alpha$ are negative and thus the
A$_{\rm v}^{{\rm H}\alpha/{\it Spitzer}}$ can not be formally
calculated and we report a 1-$\sigma$ upper limit as cB58's A$_{\rm
  v}^{{\rm H}\alpha/{\it Spitzer}}$. Given the general agreement
between A$_{\rm v}^{{\rm  H}\alpha/{\it Spitzer}}$ and A$_{\rm
  v}^{{\rm IR}}$ in a majority  of objects with secure measurements of
Pa$\alpha$ and Br$\alpha$, we adopt an average value of A$_{\rm
  v}^{{\rm H}\alpha/{\it Spitzer}}$ and A$_{\rm v}^{{\rm IR}}$ as a
representative value of \Av\ (except for the cB58 and the Clone for
which we adopt the \AvIR\ value).

\begin{figure}
\figurenum{7}
\epsscale{1.15}
\plotone{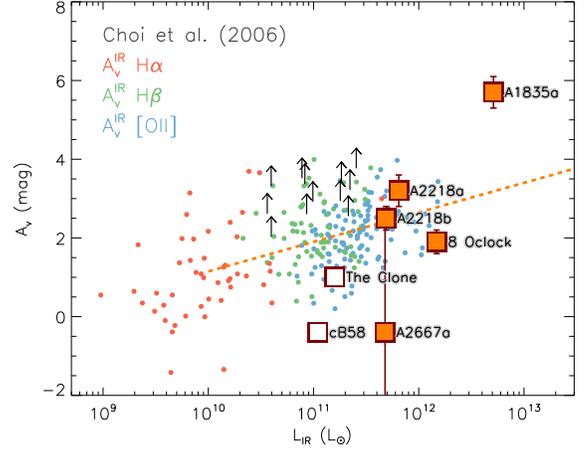}
\caption{Optical extinction, \Av, from this work compared to the
  distribution of \citet{Choi06} from the {\it Spitzer} First-Look
  Survey (FLS) with mean redshift of $z = 0.8$. Our points are shown
  in filled squares, except for the cB58 point (empty square), where we
  adopt the \AvIR\ value as an extinction estimate (see Section
  \ref{sec:result_ext}). The \citet{Choi06} points are \AvIR\ measured 
  by comparing the IR SFR, which is taken as a fiducial value, with
  the optical SFR measured from the H$\alpha$, H$\beta$, and [OII]
  emission lines. These are shown in red, blue, and grey circles,
  respectively. The intrinsic spread of points towards higher
  extinction of the Choi et al. sample may be larger if the line
  non-detections (upward arrows), which are likely due to large
  extinction, could have been measured. The intrinsic spread of
  extinction at high-$z$ is far wider than previously known from
  optical-based measurements.}   
\label{fig_Choi}
\end{figure}

\begin{figure*}
\figurenum{8}
\epsscale{1.05}
\plottwo{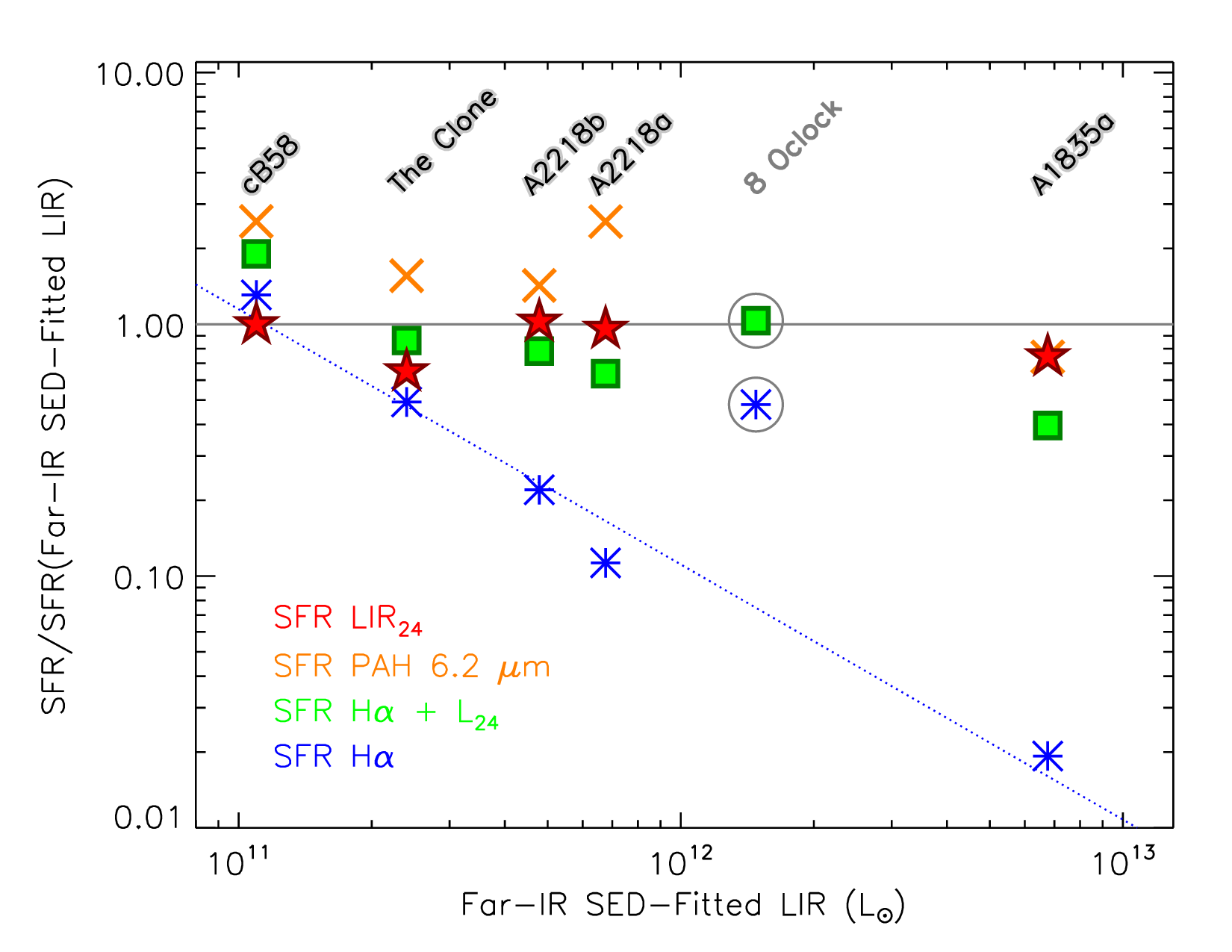}{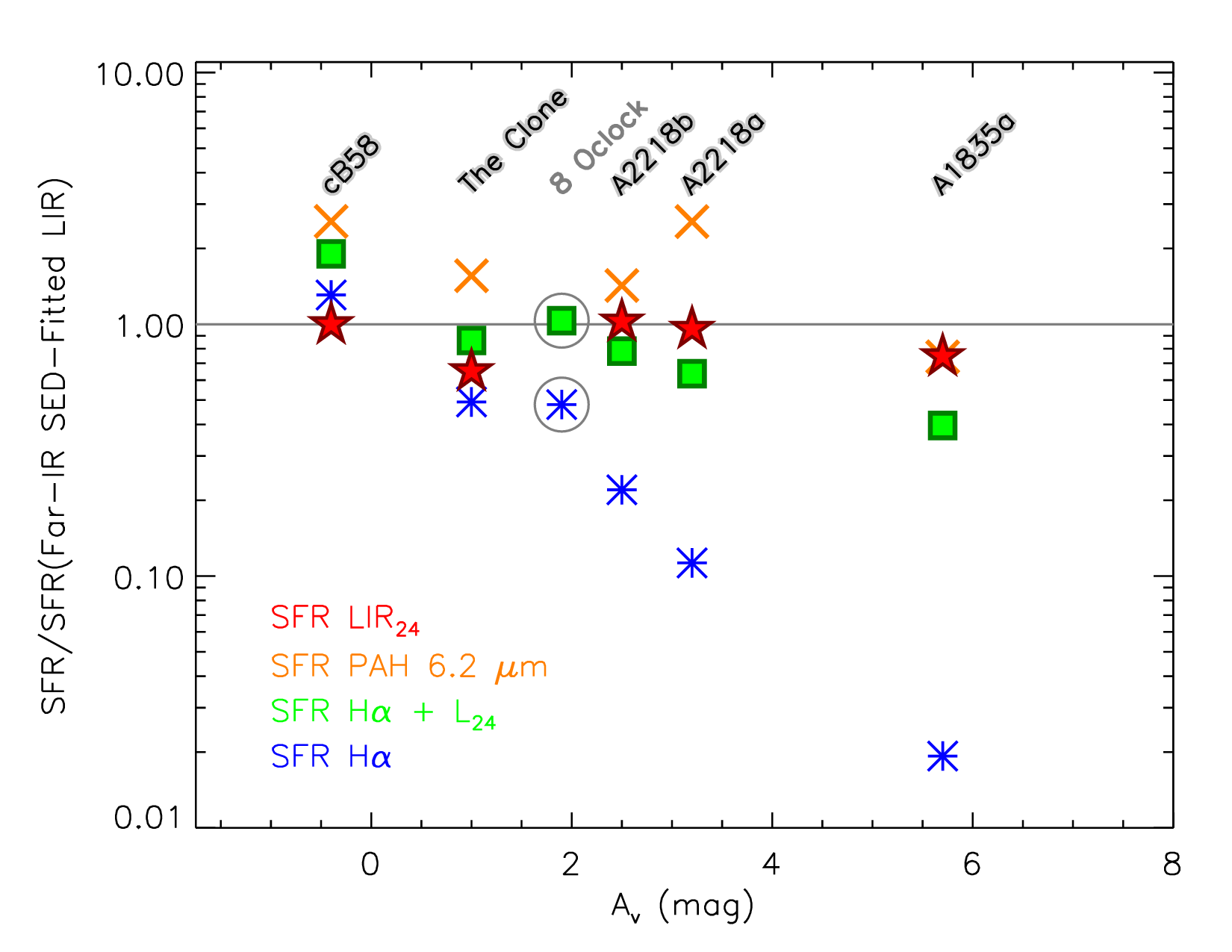}
\caption{Comparison of star formation rate (SFR) estimated from the
  uncorrected H$\alpha$, H$\alpha$ + rest-frame \LMIPS, \LTIR\ from 24
  $\micron$ indicator, and \LTIR\ from the 6.2 $\micron$ PAH emission
  line, to 
  the SFR estimated from far-IR SED fitted \LTIR, which is taken as a
  fiducial value, as a function of \LTIR\ and \Av: shown in the
  left and right panels, respectively. The H$\alpha$ SFR indicator
  systematically underestimates \LTIR\ with increasing discrepancy
  from the fiducial at larger \LTIR, while the rest are consistent
  to the fiducial within the range of uncertainties. A trend line 
  for the extinction effect on the H$\alpha$ SFR indicator in objects
  with a screen-like extinction scenario (Section \ref{sec:discuss})
  is shown as the blue dotted line. The points representing the 8
  O'clock arc are circled in grey to indicate that \LTIR\ of the arc
  is estimated from 24 $\micron$ because the SED-fitted \LTIR\ is not
  available. The deviation of H$\alpha$ SFR of the arc from the
  trend line in the left panel, however, is not due to the
  \LTIR\ measurement, but rather the differences of dust distribution 
  scenarios between the inhomogeneous mixture in the arc and the
  homogeneous distribution in the rest of the sample (see Section
  \ref{sec:result_ext})}.
\label{fig_SFRcompare}
\end{figure*}

We present our final \Av\ estimates as a function of \LTIR\ from the
24 $\micron$ indicator in Figure \ref{fig_Choi}. Again, the 24
$\micron$ \LTIR\ is used here because Abell 2667a and the 8 O'clock
arc lack far-IR \LTIR\ estimates. Our measurements are compared
with the \AvIR\ measurements by \citet{Choi06}, who study the
extinctions of objects selected in the near-IR and mid-IR from  the
{\it Spitzer} First Look Survey (FLS), which have a mean redshift of
$z = 0.8$ and luminosities ranging from the sub-LIRG to ULIRG
range. Their extinction measurements are obtained by comparing the SFR
estimates from the H$\alpha$, H$\beta$, and [OII] line fluxes with
those from IR luminosities. That is, they estimated the extinction
using the ratio of SFR from the individual optical indicators to the
IR SFR estimates, e.g. SFR$_{{\rm H}\alpha}$/SFR$_{\rm IR}$,
SFR$_{{\rm H}\beta}$/SFR$_{\rm IR}$, etc. The Choi et al. distribution
of extinction values shows a correlation between \AvIR\ and
\LTIR\ with a formal fit of $\AvIR\ = 0.75\times$log\LTIR$-6.35$,
shown as the orange line in Figure \ref{fig_Choi}. The scatter is
$\sim0.8$ mag; however, this value does not include cases where [OII]
is undetected (shown as lower limits in Figure
\ref{fig_Choi}). Allowing for these limits, the scatter is $\gtrsim1$
mag. This fit and the large scatter are similar to the result of
\citet{Afonso03}, who estimate extinction by comparing H$\alpha$ and
      [OII] SFR estimates to those from the 1.4 GHz radio continuum at
      a similar mean redshift of $z = 0.8$. A number 
of authors \citep[e.g.,][]{Bai07, Kocevski11} find that the variations
in extinction are so large that [OII] can be difficult to detect in
some galaxies that have high SFRs indicated at 24 $\micron$, although
with high quality spectra there are relatively few cases where [OII]
is completely absent \citep{Weiner07}.

We found the spread of our extinction values to be at least as
large. Abell 2218a, Abell 2218b, and the 8 O'clock arc lie within the
uncertainties of the Choi et al. relationship. Abell 2667a and cB58,
with nearly zero mag extinction are less obscured for their \LTIR,
whereas the extinction of Abell 1835a is far above the
relationship. Selection effects may have reduced the scatter in many
previous studies (e.g., the need to have an [OII] line of measurable
strength). Our results indicate that galaxies up to $z \sim 3$ show a
very large range of \Av, as is the case in the local Universe.

\section{Effects of Extinction on Star Formation Rate
  Indicators}\label{sec:discuss} 

In Figure \ref{fig_SFRcompare}, we compare the SFR estimates from
H$\alpha$ (uncorrected for dust extinction), H$\alpha$ + \LMIPS,
\LMIPS, and the 6.2 $\micron$ PAH emission with those of
\LTIR\ measured by far-IR SED fitting as a function of \LTIR\ (Figure
\ref{fig_SFRcompare} left) and \Av\ (Figure \ref{fig_SFRcompare}
right). Abell 2667a is excluded from this comparison due to its lack
of the far-IR SED fitted \LTIR\ and the fact that the curvature of the
lensed arc caused a fraction of H$\alpha$ flux to fall off the slit
(the position of the slit is shown 
in the bottom left panel of Figure \ref{fig_slitpos}; a part of the
`A2' arc is outside the slit). Without the correction for dust
extinction, the H$\alpha$ SFR indicator is clearly affected at high
\LTIR\ and extinction. For the cases that extinction behavior
resembles a uniform dust screen (every galaxy except the 8 O'clock
arc), there is a tight trend of the H$\alpha$ SFR deviation from the
fiducial SED-fitted SFR as a function of the \LTIR\ that is given by 
\begin{equation} 
\label{eq3}
 {\rm log}[{\rm SFR}({\rm H}\alpha) / {\rm SFR}({\rm LIR})] = 11.21 -
 1.01~{\rm log(LIR)}
\end{equation}

Applying a single overall extinction to H$\alpha$ will 
increase the estimates of SFRs and improve the agreement to the
\LTIR\ SFR, but will not correct the trend with \LTIR. Further
improvement in H$\alpha$-based SFRs require introducing corrections
for extinction as a function of stellar mass \citep[e.g.,
][]{Moustakas06, Weiner07}, or of the SFR itself
\citep[e.g.][]{Buat05}; see Section \ref{sec:discuss}. 

The deviation of the H$\alpha$ point of 8 O'clock arc in Figure
\ref{fig_SFRcompare} (left) indicates that a larger amount of
H$\alpha$ flux is escaping from the arc given its \LTIR, which could
be due to a different dust distribution relative to the star forming
regions. The indicator is usually applied with a single nominal level
of extinction; in this case, the line in Figure \ref{fig_SFRcompare}
(left) would be shifted upward but the slope would not be corrected. 

For Abell 2218a, Abell 1835a, and the 8 O'clock arc where Pa$\alpha$
line is well-detected, we apply the extinction correction to the
Pa$\alpha$ line luminosity and compare the ratio of
\LMIPS/$L$(Pa$\alpha_{\rm corr}$) with the relationship that
\citet{AH06} found for local LIRGs. \citet{Papovich09} made this
comparison for Abell 2218a and found the \LMIPS/$L$(Pa$\alpha_{\rm
  corr}$) ratio to be $\simeq 0.5$ dex lower than the local relation,
indicating that Abell 2218a has lower \LMIPS\ than local galaxies of
comparable $L({\rm Pa}\alpha_{\rm corr})$, while agreeing with those
of local individual HII regions. Papovich et al. interpreted this
difference as an indication that Abell 2218a harbors extended
star-forming regions similar to a scaled-up local HII regions, rather
than nuclear starburst like local U/LIRGs. In our analysis, we have
reproduced the \LMIPS/$L$(Pa$\alpha_{\rm corr}$) ratio measured by
Papovich et al. using our reduction technique for Abell 2218a,
log[\LMIPS/$L$(Pa$\alpha_{\rm corr}$)] = $2.2 \pm 0.2$, and also found
that the ratio for Abell 1835a and the 8 O'clock arc are $0.9 - 1.1$
dex lower than the \citet{AH06} relationship. This is consistent with
the result in Table \ref{table_derivedqn} that the
\citet{Rujopakarn12} single-band 24 $\micron$ \LTIR\ indicator yields
\LTIR\ values agreeing with the far-IR SED-fitted \LTIR, which implies
that these galaxies have extended star formation (see Section
\ref{sec:result_lir} for details of the indicator). That is, the local
relationship of $L$(Pa$\alpha_{\rm corr}$) and \LMIPS\ reported by
\citet{AH06} could have limited applicability at high redshift because the 
Pa$\alpha$ line will systematically have lower extinction for a given
\LMIPS\ due to the extended structure of star formation.

All the other indicators give consistent estimates, within the
expected errors of $\sim 0.2$ dex. We found that the H$\alpha$ +
\LMIPS\ indicator by \citet{Kennicutt09} tends toward a smaller value
of SFR at large extinctions or large \LTIR, particularly above
$10^{12}$ \Lsun, as observed in Abell 1835a and the 8 O'clock arc. The
SFR estimates from the 6.2 $\micron$ PAH emission do not show a
systematic trend with \LTIR, although the scatter is larger than with
other indicators. The scatter for individual galaxies is $\sim$ 0.2
dex, similar to the scatter of \LTIR\ values we have found in
\citet{Rujopakarn12} at redshifts where the 24 $\micron$ band probes
the PAH emissions. Since these are high S/N measurements of an
individual (i.e. relatively non-blended) PAH emission line, this
result suggests that the \citet{Rujopakarn12} SFR indicator has
succeeded in correcting for the SED evolution.

\section{CONCLUSIONS}\label{sec:conclusions}

We observed four strongly gravitationally lensed star-forming galaxies
at $1 < z < 3$ with {\it Spitzer}/IRS and the LBT/LUCIFER to obtain
the mid-IR and near-IR spectroscopy. These observations are targeted
to cover IR recombination lines, including H$\alpha$ in the
near-IR and Pa$\alpha$ or Br$\alpha$ in the mid-IR. We include another
three galaxies from the literature with similar suites of
observations, yielding a total sample of seven galaxies. Our sample
spans the redshift range of $1.03 - 2.73$ and the \LTIR\ range of $1.3
\times 10^{11}$ \Lsun\ to $7.0 \times 10^{12}$ \Lsun.

The IR recombination line ratios are used to measure extinction that 
can probe deep into the highly obscured star-forming regions and thus
provide an unbiased measure of extinction under the foreground
screen assumption. Independently, we estimate the extinction by
comparing the optical and IR SFRs, a method that does not make the
foreground screen assumption. The results from the two methods are
consistent in three out of four galaxies with good IR recombination
line flux measurements, suggesting that the dust extinction in these
galaxies is consistent with a foreground screen (i.e. uniform dust
distribution). However, in the fourth case, the extinction estimates
from two methods disagree by 1.8 mag, indicating a deviation from the
uniform dust screen assumption, which suggests an inhomogeneous dust
mixture. The extinction range of our sample (assuming a foreground
screen) is $\sim 0.0 - 5.9$ mag, which is a larger spread than
previously known for intermediate and high redshift galaxies based on
measurements with optical emission lines. These results suggest a
large diversity in both the extinction levels and dust distribution
scenarios at high redshift.

We compare the performance of various SFR indicators over the
extinction range and find that substantial extinction corrections are
required for the H$\alpha$-based SFR indicator. The remaining
indicators (1) combined H$\alpha$ and \LMIPS; (2) \LMIPS; and (3) PAH
(6.2 $\micron$) all give estimates consistent to within the expected
uncertainties of $\sim0.2$ dex.

We thank Brian Siana and Eiichi Egami for helpful discussions. WR
thanks Alexandra Pope and Philip Choi for data points in Figures
\ref{fig_Pope} and \ref{fig_Choi}, respectively, and acknowledges the
support from the Thai Government Scholarship. This work is supported
by contract 1255094 from Caltech/JPL to the University of Arizona.\\

\end{document}